\documentclass[reqno,11pt]{article}
\RequirePackage{amsthm, amsmath, amsfonts, natbib, graphicx}
\usepackage{supertabular}
\usepackage{bm}
\usepackage{booktabs}
\usepackage{multirow}
\usepackage{array}
\usepackage{subfig}
\usepackage{xcolor}

\usepackage{algpseudocode}

\usepackage{caption}
\algnewcommand{\LineComment}[1]{\State \(\triangleright\) #1}
\long\def\omitthis#1{}
\usepackage{amsmath,amsfonts,mathtools}
\usepackage{bbm}

\newcommand{\btheta}{\boldsymbol{\theta}}

\newcommand{\R}{\mathbb{R}}

\newcommand{\mU}{\mathcal{U}}

\newcommand{\E}{\mathbb{E}}

\newcommand{\bh}{{\boldsymbol{h}}}

\newcommand{\be}{{\boldsymbol{e}}}

\newcommand{\bw}{{\boldsymbol{w}}}
\newcommand{\bx}{{\boldsymbol{x}}}

\newcommand{\bz}{{\boldsymbol{z}}}

\newcommand{\bG}{{\boldsymbol{G}}}
\newcommand{\bD}{{\boldsymbol{D}}}

\newcommand{\bC}{{\boldsymbol{C}}}

\newcommand{\BT}{\boldsymbol{\theta}}
\newcommand{\BD}{\boldsymbol{\delta}}

\newlength\myindent
\newcommand{\argmin}{\operatornamewithlimits{argmin}}
\newcommand{\argmax}{\operatornamewithlimits{argmax}}

\usepackage{amsmath,amsfonts,graphicx,amsthm,hyperref}
\usepackage{float}
\usepackage[ruled]{algorithm}

\title{A Review of Adversarial Attack and Defense for Classification Methods}
\author{Yao Li \footnote{{\tt yaoli@unc.edu}, Department of Statistics \& Operations Research, University of North Carolina at Chapel Hill}
\and Minhao Cheng \footnote{{\tt mhcheng@g.ucla.edu}, Department of Computer Science, University of California at Los Angeles}
\and Cho-Jui Hsieh \footnote{{\tt chohsieh@cs.ucla.edu}, Department of Computer Science, University of California at Los Angeles}
\and Thomas C. M. Lee\footnote{{\tt tcmlee@ucdavis.edu}, Department of Statistics, University of California at Davis}
}
\date{August 18, 2021; revised: November 5, 2021}

\begin{document}        
\maketitle
\begin{abstract}
Despite the efficiency and scalability of machine learning systems, recent studies have demonstrated that many classification methods, especially deep neural networks (DNNs), are vulnerable to adversarial examples; i.e., examples that are carefully crafted to fool a well-trained classification model while being indistinguishable from natural data to human. This makes it potentially unsafe to apply DNNs or related methods in security-critical areas.  Since this issue was first identified by \cite{biggio2013evasion} and \cite{szegedy2013intriguing}, much work has been done in this field, including the development of attack methods to generate adversarial examples and the construction of defense techniques to guard against such examples.  This paper aims to introduce this topic and its latest developments to the statistical community, {\color{black}primarily focusing on the generation and guarding of adversarial examples}. Computing codes (in python and R) used in the numerical experiments are publicly available for readers to explore the surveyed methods. It is the hope of the authors that this paper will encourage more statisticians to work on {\color{black} this important and exciting field of generating and defending against adversarial examples.}



\end{abstract}
{\bf Keywords:} adversarial examples, adversarial training, deep neural networks, defense robustness


\newpage
\section{Introduction}
\label{sec:intro}

\begin{figure}[H]
    \centering
    \includegraphics[width=0.8\textwidth]{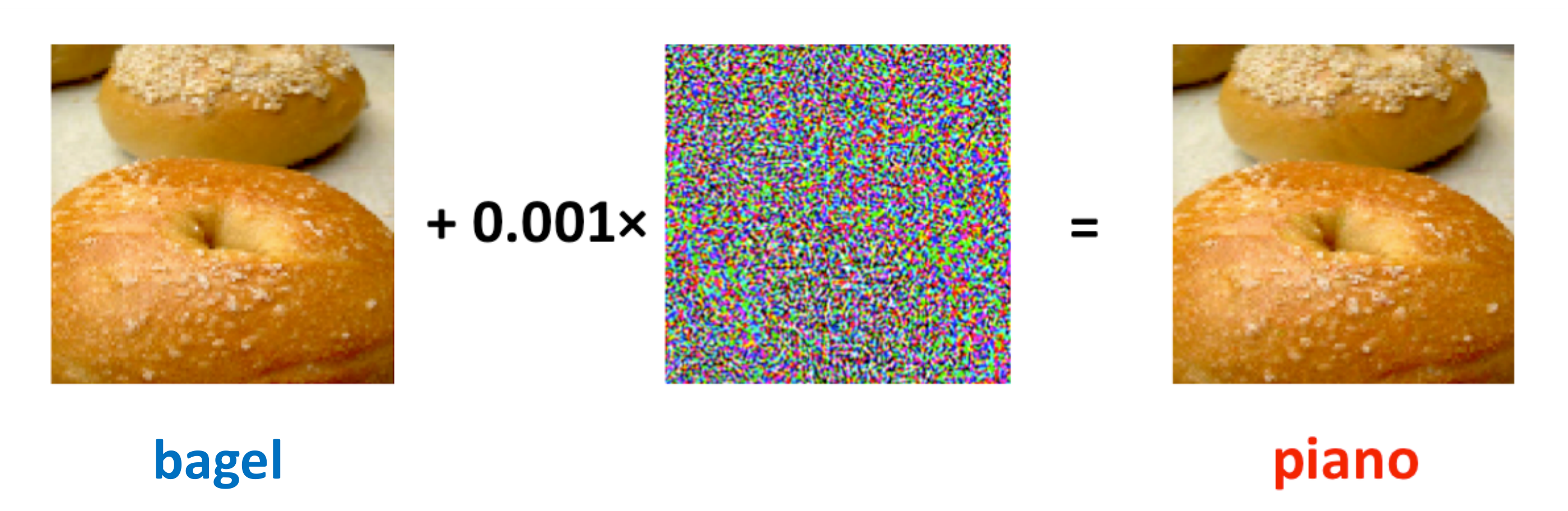}
    \caption{An example of adversarial example, taken from~\cite{chen2017zoo}.}
    \label{fig:adv_example}
\end{figure}

Machine learning systems achieve state-of-the-art performance in various tasks in artificial intelligence, such as image classification, speech recognition, machine translation and game-playing~\citep[e.g.,][]{simonyan2014very,silver2016mastering,devlin2018bert}. Despite their tremendous successes, machine learning models have been shown to be vulnerable to adversarial examples. By adding imperceptible perturbations to the original inputs, the attacker can produce adversarial examples to fool a learned classifier~\citep[e.g.,][]{szegedy2013intriguing,goodfellow2014explaining}. {\color{black}Adversarial examples are indistinguishable from the original inputs to human, but are mis-classified by the classifier. For an illustration, consider the following images in Figure~\ref{fig:adv_example}:}

To humans, the two images appear to be the same -- the vision system of human will identify each image as a bagel. The image on the left side is an ordinary image of a bagel (the original image). However, the image on the right side is generated by adding a small and imperceptible perturbation that forces a particular classifier to classify it as a piano. {\color{black} The example in Figure~\ref{fig:adv_example} was generated to fool a deep neural network (DNN) based classifier. However,} an adversarial example is not an isolated case that happens only to some neural networks, but a general case for most of the practically used neural networks.
Besides, recent works have shown that linear models and many other general statistical learning models, such as logistic regression and tree-based models, are also vulnerable to adversarial examples~\citep{chen2019robustness}. With the wide application of machine learning models, this causes serious concerns about the safety of machine learning systems in security sensitive areas, such as self-driving cars, flight control systems, healthcare systems and so on. Therefore, studies on adversarial examples have attracted much attention in recent years. 

Most of the studies on adversarial examples can be generally classified into two categories: works on how to efficiently generate adversarial examples (i.e., attack), and works on how to guard against such adversarial examples (i.e., defense). This paper reviews the latest research on adversarial attack and defense, and compares the state-of-the-art attack and defense methods on benchmark datasets.
We illustrate the main concepts via an image classification task with a continuous input space. However, we also note that there have been research activities on adversarial robustness for other applications with discrete input data, such as natural language processing (NLP), discrete models and nearest neighbor classifiers~\citep[e.g.,][]{jia-liang-2017-adversarial,samanta2017towards,gao2018black,cheng2020seq2sick,JMLR:v21:19-569}. 

With this paper, it is our hope that more statisticians will be engaged in this important and exciting field of attacking and defending classification methods.  As to be demonstrated below, many problems from this field are statistical in nature, and statistical methodologies and principles can be adopted to improve existing methods or even to develop new methods.
R codes that demonstrate how to attack a logistic regression model are provided for statisticians who are interested in this problem to experiment with.  {\color{black} Lastly, we note that there are other review papers in related topics that might be of interest to the reader; e.g., \cite{serban2020adversarial} reviewed adversarial phenomenon in the field of object detection, while other authors~\citep[e.g.,][]{qiu2019review,yuan2019adversarial,ren2020adversarial,xu2020adversarial} provided reviews of adversarial examples with emphasis in computer vision and natural language processing from the perspective of computer science.  These authors categorized the methods in a different manner when compared to ours.}

\textbf{Notation.} In this paper, all the vectors are denoted as bold symbols. The input to the classifier is represented by $\bx$ and the label associated with the input is represented by $y$. Thus, one observation is a pair $(\bx,y)$. The classifier is denoted as $f(\cdot)$ and $f(\bx)$ represents the output vector of the classifier. The dimension of $f(\bx)$ is equal to the number of classes of the dataset. Let $f(\bx)_i$ denote the score of predicting $\bx$ with label $i$.  We caution that, in some works in the literature, $f(\bx)_i$ is taken as the ``probability'' that sample $\bx$ corresponds to label $i$, but very often these $f(\bx)_i$'s are not probabilities. The prediction of the classifier is denoted as $c(\bx)=\argmax\limits_if(\bx)_i$; that is, the predicted label is the one with the highest prediction score. We use the $\ell_\infty$ and $\ell_2$ distortion metrics to measure similarity between inputs and report the $\ell_\infty$ distance in the normalized $[0,1]$ space, and the $\ell_2$ distance as the total root-mean-square distortion normalized by the total number of dimensions. 


The rest of this paper is organized as follows.  Section~\ref{sec:dnn} provides a quick summary of DNNs.  This section can be omitted if the reader has a good understanding of the classification task, by treating a DNN as a black-box classifer.  Then Section~\ref{sec:attack} describes the latest developments for attacking classification methods, while Section~\ref{sec:defense} presents defense methods for defending against such attacks.  Lastly, numerical comparisons are given in Section~\ref{sec:exp} while concluding remarks, including future directions for statistical research, are offered in Section~\ref{sec:conclude}.

\section{Deep Neural Networks}
\label{sec:dnn}
Deep Neural Networks (DNNs) have become one of  the most prominent technologies of our time, as they achieve state-of-the-art performance in many machine learning tasks, including but not limited to image classification, text mining, and speech processing. A brief introduction to DNNs is given in this section to help readers understand attack and defense methods. More details of DNNs can be found, for example, in~\cite{Goodfellow-et-al-2016}. {\color{black} To speed up the pace, the readers may skip this section (and come back later), which will not affect understanding the concepts behind those attack and defense methods. It is because the DNNs described below can be viewed as black-box classifiers and their details are not crucial for the comprehension of such concepts. However, for readers who are interested in the robustness of DNNs and not familiar with their architectures, this section would be helpful.}

\begin{figure}
    \centering
    \includegraphics[width=0.8\textwidth]{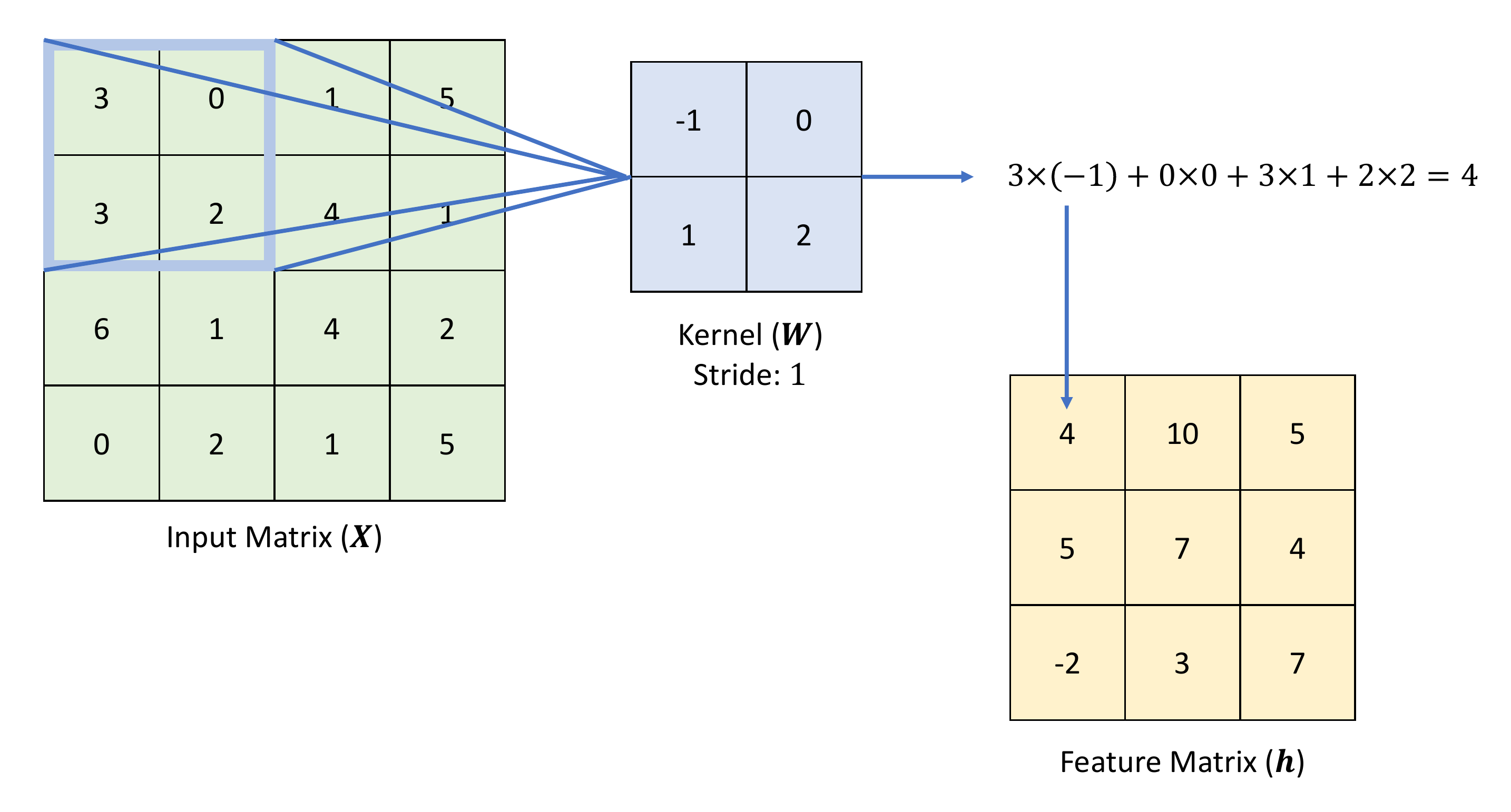}
    \caption{Convolutional operation with stride (step size) equals to 1.}
    \label{fig:conv}
\end{figure}

A DNN consists of a large class of models and learning methods. Here we describe a general form of DNN classifier that is widely applied in image classification. For a DNN model $f(\cdot)$ with $L$ layers, each layer is parameterized by a weight matrix $\BT_i$, which holds the knowledge of the DNN model and is updated during the training process. Each neuron of the DNN has an activation function $\phi_i$, which is usually used to introduce non-linearity into the output of a neuron. Each function $\phi_i$ for $i\in\{1,...,L\}$ is modeled using a layer of neurons, where the function $\phi_i$ takes the output of the previous layer ($\bh_{i-1}$) as input and produces $\bh_i$ as the output of the current layer. The function $\phi_i$ is usually chosen to be ReLU (rectified linear unit) $\phi(v)=\max(0,v)$, 
but there are also many other choices. In fully connected neural networks, the output of each layer is generated by multiplying a weight matrix with the previous layer output and adding a bias term. However, there could be other designs for the weights. For example, in a convolutional neural network (CNN), the weight $\BT_i$ is a kernel matrix. 
The kernel matrix scans over the input matrix, from left to right and from top to bottom, and each one of the values within the kernel is multiplied by the input value on the same position to generate the output matrix; see Figure~\ref{fig:conv} for an example. More details of convolutional neural network design can be found in~\cite{lecun1998gradient}.
In general, a DNN can be expressed as:
\begin{align*}
    f(\bx)=\phi_L(\BT_L,\phi_{L-1}(\BT_{L-1},...\phi_2(\BT_2,\phi_1(\BT_1,\bx)))).
\end{align*}

\begin{figure}
    \centering
    \includegraphics[width=\textwidth]{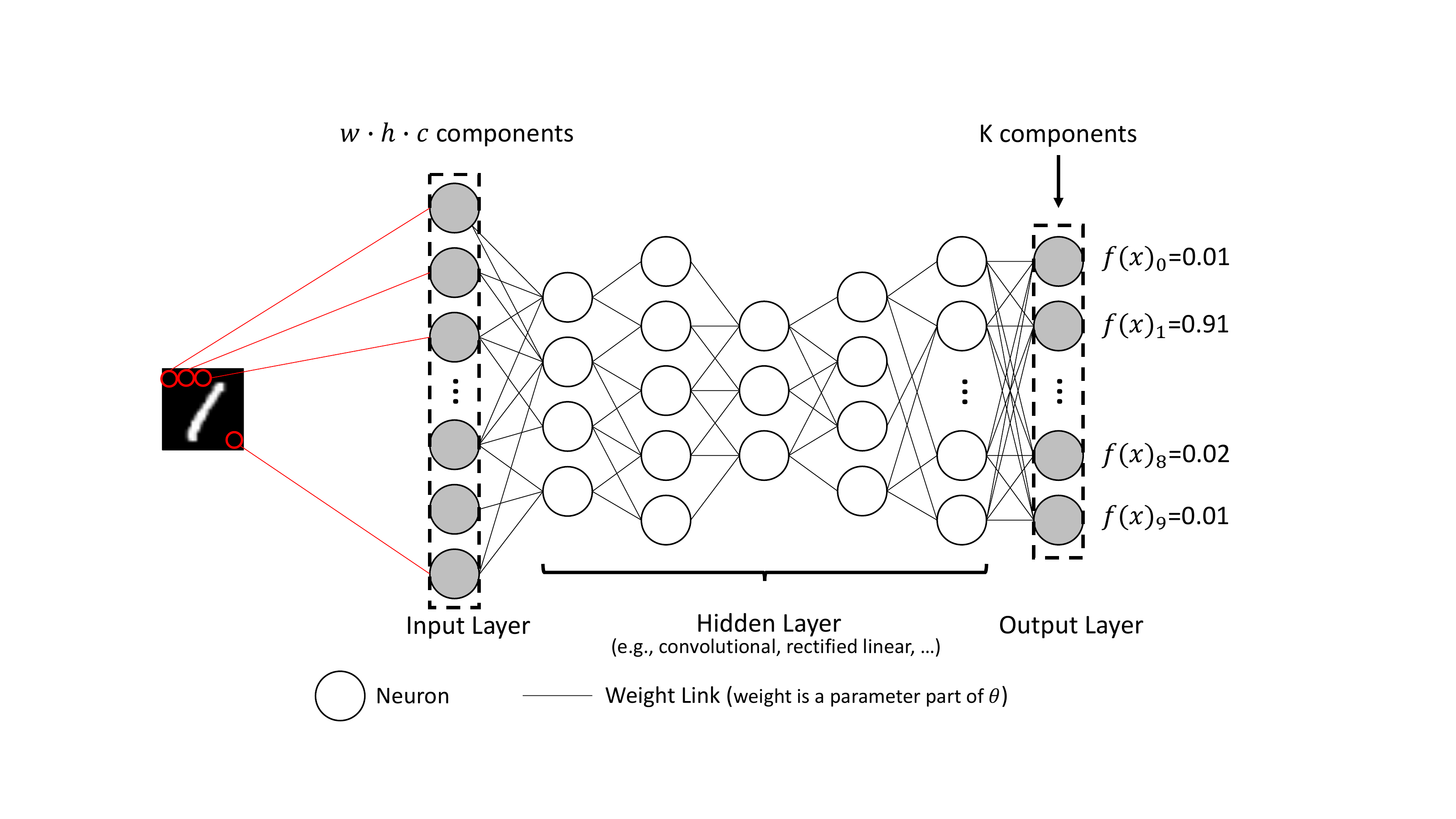}
    \caption{{\bf DNN Classifier:} the model takes an image of a handwritten digit as input and predicts the probability of it being in one of the $K=10$ classes for digits $0$ to $9$~\citep{papernot2017practical}.}
    \label{fig:dnn}
\end{figure}

As illustrated in Figure~\ref{fig:dnn}, a deep classifier is a machine learning model that uses a hierarchical composition of $L$ parametric functions to model an input $\bx$. 
Depending on the loss applied to the final layer, a neural network can easily handle both regression and classification tasks. 
For regression, there is only one output unit at the end of the network. For multi-class classification, there are $K$ output units, where $K$ is the number of classes. 

For the classification task, the cross-entropy loss is widely used:
\begin{align*}
    \ell(\bx,y)=-\sum_{i=1}^K\mathbbm{1}_{\{i=y\}}\log(f(\bx)_i). 
\end{align*}
Given a dataset with many known input-label pairs $(\bx,y)$, the DNN will adjust its parameters ($\{\BT_1,...,\BT_n\}$) to reduce the prediction error between the prediction and the correct label. 
The generic approach to minimize the prediction loss is gradient descent, called back-propagation in this setting~\citep{hecht1992theory}. Because of the compositional form of the model, the error gradient is derived using the chain rule for differentiation. 
The gradient is computed by a two-pass algorithm, where a forward pass is used to compute the loss and a backward pass is carried out to derive the gradient.
In backward pass, error gradients with respect to network parameters are successively propagated from the network's output layer to its input layer. For large datasets, stochastic gradient descent is usually applied, where the gradient descent is done on a random subset of data.

At the testing stage, the DNN with fixed parameters is used to predict the label of unseen input. The output of the DNN is a probability vector $f(\bx)$, and the prediction is 
$c(\bx)=\argmax\limits_if(\bx)_i$. 

DNNs achieve state-of-the-art performance in the task of image classification.
{\color{black} For example, an vision transformer network~\citep{dosovitskiy2021image} can achieve over $99\%$ accuracy when classifying natural images from the benchmark dataset CIFAR10~\citep{krizhevsky2009learning}.} 
Ideally, we expect the deep classifier to generalize well, making accurate predictions for inputs outside of the domain explored during training. However, recent studies on adversarial attack show that carefully crafted adversarial examples can easily fool the model.

\section{Attack}
\label{sec:attack}


\begin{figure}
    \centering
    \includegraphics[width=0.8\textwidth]{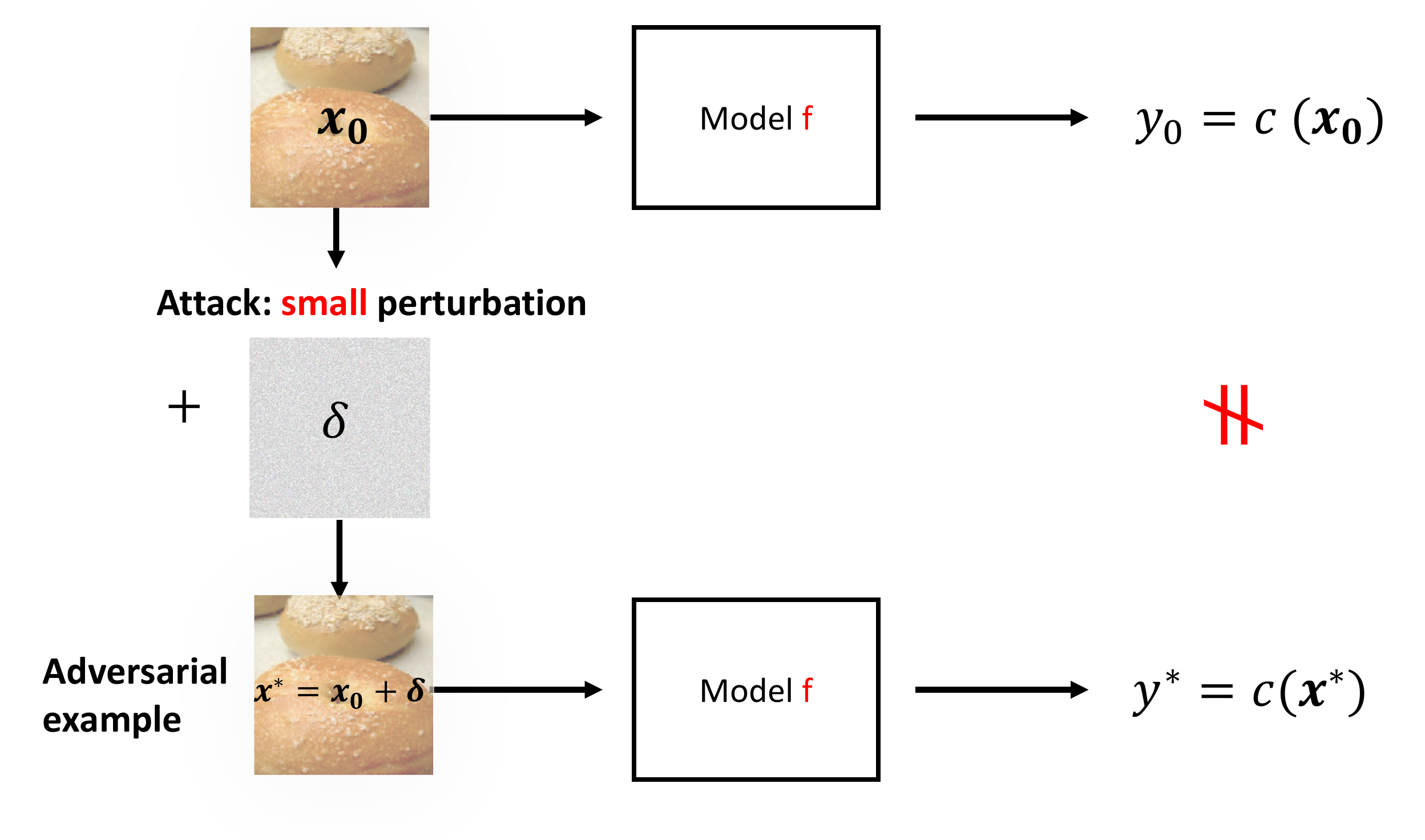}
    \caption{Illustration of adversarial attacks.}
    \label{fig:notation}
\end{figure}

Adversarial attacks are processes to generate an adversarial example based on a given natural sample and the victim model. 
Figure~\ref{fig:notation} illustrates this process of generating adversarial examples. Here $\bx_0$ denotes the natural input, and the DNN can correctly predict its label $y_0$.  An adversarial attack aims to find a small perturbation $\boldsymbol{\delta}$ such that the adversarial example $\bx^* = \bx_0 + \boldsymbol{\delta}$, which looks similar to $\bx_0$ to humans, will be mis-classified by the victim model.  

Multiple attack methods have been introduced for the crafting of adversarial examples to attack various DNNs. In general, there are two kinds of attack goals: targeted and untargeted. For untargeted attack, the attack is successful if the input is predicted with any wrong label. Take Figure~\ref{fig:adv_example} as example. As long as the bagel image is not classified as bagel, the attack is successful. For targeted attack, the attack is successful only when the adversarial example is classified as the target class. In this example, if the target class is piano, the attack is successful only when the right side image is labeled as piano by the victim classifier. 

Based on the information needed, the attack methods can be grouped into three categories: (1) gradient-based, (2) score-based, and (3) decision-based. Most of these methods can perform both targeted and untargeted attacks. In general, one attack method belongs to one of the three categories, while recently it has been shown that an ensemble of attacks from multiple categories can potentially lead to a stronger attack~\citep{croce2020reliable}. 
{\color{black} If all the information of the victim model, such as model structure, parameters and so on, is revealed to the attacker, the scenario is called white box setting.} If only the predicted scores are available, the scenario is called soft-label black box setting. If only the predicted labels are revealed, the scenario is called hard-label black box setting. There are also gray box settings, where part of the model information is available.  {\color{black}Notice that the part of the model information that is available/unavailable is problem dependent.  For example, \cite{yang2020ml} defined the gray box as the scenario where the attacker has access to the classifier but not the design of the adversarial example detector. We will thus focus on white-box and black-box attacks in this paper.}
The next three subsections present representative examples and deeper discussions of the three categories of attack methods.


\subsection{Gradient-Based Attack}

Many existing attack methods fall in this category. These methods leverage the gradients of the loss w.r.t. the input to form adversarial examples. 
For instance, the Fast Gradient Sign Method (FGSM)~\citep{goodfellow2014explaining} generates adversarial examples based on the sign of gradients, and uses a step size to control the $\ell_\infty$ norm of perturbation. The Basic Iterative Method (BIM)~\citep{kurakin2016adversarial} and Projected Gradient Descent Attack (PGD)~\citep{madry2017towards} can be viewed as iterative versions of FGSM. PGD crafts adversarial examples by running FGSM for a fixed number of iterations or until mis-classification is achieved. The most effective gradient-based adversarial attack methods to date are C$\&$W~\citep{carlini2017towards} and PGD. Both C$\&$W attack and PGD have been frequently used to benchmark the defense algorithms due to their effectiveness~\citep{athalye2018obfuscated}.
Since gradient information is required to perform attack, gradient-based attack is mainly for the white-box setting. 
The process of crafting adversarial examples can be formulated as an optimization problem. Depending on the optimization formulations of the attack methods, gradient-based attack methods can be further divided into two sub-categories.

{\bf Constraint-optimization formulation based methods}: The first sub-category consists of constraint-optimization formulation based methods. Given a model with a fixed parameter $\BT$ and an input pair $(\bx_0, y_0)$, the process of generating adversarial examples $\bx^*$ can be described by the following optimization problem:
\begin{equation}
\bx^* = \bx_0 + \boldsymbol{\delta} \quad \mbox{with} \quad
\boldsymbol{\delta} = \arg\max_{\boldsymbol{\delta}\in\mathcal{S}}L(\BT,\bx_0+\boldsymbol{\delta},y_0),
\label{eq:attack_cate_1} 
\end{equation}
where $L$ is the loss function, $\boldsymbol{\delta}$ is the adversarial perturbation, and $\mathcal{S}\in\R^d$ is the set of allowed perturbations, usually chosen to be $\mathcal{S}=\{\boldsymbol{\delta}|\|\boldsymbol{\delta}\|_\infty\leq \epsilon\}$ for some $\epsilon$.


{\color{black} The optimization goal is to search for the adversarial perturbation that leads to wrong classification. Therefore, for untargeted attack, the loss function $L$ can be the loss function used to train the classifier, while for targeted attack, other loss functions designed for the target class should be used. By maximizing~(\ref{eq:attack_cate_1}), the attacker is forcing the classifier to err on the classification task. To guarantee that the perturbation is not too large so that $\bx_0$ and $\bx^*$ are indistinguishable to human, the search space is restricted to be within the $\epsilon$-ball around the input.} 

The optimization problem defined in~(\ref{eq:attack_cate_1}) can be reformulated as
\begin{equation}
\color{black} \arg\max_{\BD\in\mathcal{S}}L(\boldsymbol{\theta},\bx_0+\BD,y_0)  =\argmax_{\|\BD\|_\infty\le\epsilon}L(\boldsymbol{\theta},\bx_0+\BD,y_0),
\label{eq:pgd_op}  
\end{equation}
which can be solved by gradient descent. {\bf Projected-Gradient Descent Attack (PGD)} crafts adversarial examples by solving this optimization problem with the projected gradient descent method. It is proposed by~\cite{madry2017towards}, which finds adversarial examples in an $\epsilon$-ball of the input. The PGD attack updates in the direction that decreases the probability of the original class most, then projects the result back to the $\epsilon$-ball of the input.  The updating equation of the PGD attack is:
\begin{align}
 \bx^{t+1}=\Pi_\epsilon\left\{\bx^t+\alpha\cdot\text{sign}\Big(\nabla_\bx L(\boldsymbol{\theta},\bx^t,y)\Big),\bx_0\right\},    
 \label{eq:pgd}
\end{align}
where $\bx_0$ is the original input, $\bx^t$ is the updated input in step $t$, $\epsilon$ controls the maximum distortion, $\alpha$ is the step size and $\nabla_\bx L(\boldsymbol{\theta},\bx^t,y)$ represents the gradient of classification loss $L(\cdot)$ w.r.t. input $\bx^t$.
Basically, PGD attack constructs an adversarial example by adding or subtracting a small error term, $\alpha$, to each input dimension. 
The decision of adding or subtracting an error term depends on whether the sign of the gradient for an input dimension is positive or negative. 
Then the result from previous step is projected to the $\epsilon$-ball around the original input. If~(\ref{eq:pgd}) is run for only one step, it is equivalent to the Fast Gradient Sign Method (FGSM)~\citep{goodfellow2014explaining}. If it is run for multiple steps, it is the PGD attack or the Basic Iterative Method (BIM)~\citep{kurakin2016adversarial}. Recently, \cite{croce2020reliable} introduced a new attack method called Autoattack, which combines two extensions of PGD and a decision-based attack to automatically evaluate defense methods without parameter tuning.



{\bf Regularization-optimization formulation based methods}: The second sub-category consists of regularization-optimization formulation based methods. The {\bf C$\&$W attack} proposed by~\cite{carlini2017adversarial} is a representative one, and is by far one of the strongest attack methods.  It can perform both targeted and untargeted attacks, which are formulated as the following optimization problem:
\begin{align}
       \bx^*=\argmin\limits_\bx\left\{\|\bx-\bx_0\|_2^2+cg(\bx)\right\},
              \label{eq:cw_op}
\end{align}
where the first term enforces small distortion of the original input and the second term is the loss function that measures the success of the attack. The parameter $c>0$ controls the trade-off between distortion and attack success. 

If it is for untargeted attack, which means the attacker only wants the classifier to make a mistake and does not care the predicted label of the adversarial example, then $g(\bx)$ is defined as:
\begin{align*}
    g(\bx)=\max\{f(\bx)_{y_0}-\max\limits_{i\ne y_0}f(\bx)_i,0\},
\end{align*}
where $y_0$ denotes the true label of input $\bx_0$, $f(\bx)_{y_0}$ represents the score of predicting input $\bx$ with label $y_0$, and $f(\bx)_i$ represents the score of predicting input $\bx$ with label $i$. Minimizing $g(\bx)$ will make the prediction score of true class $y_0$ smaller than the prediction scores of classes other than $y_0$. Therefore, the adversarial example $\bx^*$ will be classified to a wrong class.

If the attacker wants the adversarial example to be classified into a specific target class with label $t$, where $t\ne y_0$, then $g(\bx)$ is defined as:
\begin{align*}
    g(\bx)=\max\{\max\limits_{i\ne t}f(\bx)_i-f(\bx)_t,0\}.
\end{align*}
The above targeted attack loss function will push the prediction score of the target class $t$ to be higher than those of other classes. 
\cite{carlini2017adversarial} showed that their attack can successfully bypass ten different defense methods designed for detecting adversarial examples. 

Inspired by an elastic-net method proposed by~\cite{zou2005regularization}, \cite{chen2018ead} proposed the elastic-net attack (EAD), which also belongs to this sub-category of regularization-optimization formulation based methods. EAD formulates the process of generating adversarial examples as an elastic-net regularized optimization problem, which can be viewed as extended version of the C\&W Attack. Their formulation of elastic-net attacks to DNNs for crafting an adversarial example with respect to a labeled sample is as follows:
\begin{align*}
       \bx^*=\argmin\limits_\bx\left\{\|\bx-\bx_0\|_2^2+\beta\|\bx-\bx_0\|_1+cg(\bx)\right\},
\end{align*}
where $g(\bx)$ is the same as the one defined in C\&W attack. {\color{black} The authors showed that EAD can improve attack transferability and complement adversarial training.}

We note that most gradient-based attack methods fall into the above two sub-categories, but there are also other kinds of gradient-based attack methods, such as Jacobian-based Saliency Map Attack~\citep{papernot2016limitations}, Deepfool~\citep{moosavi2016deepfool}, Network for Adversary Generation~\citep{nag-cvpr-2018} and Adversarial Transformation Networks~\citep{46527}. All of these methods perform adversarial attacks based on the gradient information.

{\bf Choice of perturbation set in adversarial attacks. } Both constraint and regularization optimization based formulations require a distance measurement to quantify the difference between the original example and the perturbed example. Despite $\ell_\infty$ norm being used for PGD-based attacks \eqref{eq:attack_cate_1} and the $\ell_2$ norm being widely used in C\&W based attacks \eqref{eq:cw_op}, in general we can use any $\ell_p$ norm for both constraint and regularization based formulations. For example, $\ell_1$ norm and a mixed of different norms have been used in~\citep{chen2018ead}. Further, recently it has been recognized that $\ell_p$ norm based attacks may not be realistic, so researchers have started to explore more flexible perturbation sets. For instance, \cite{wong2019wasserstein} uses  Wasserstein distance to measure the perturbation strength, and \cite{wong2020learning} proposed learning the perturbation sets based on human perception. 

\subsection{Score-Based Attack}

In reality, detailed model information, such as the gradient, may not be available to the attackers. The score-based attack methods do not require access to gradients. They perform adversarial attacks based on the output scores $f(\bx)_i$'s of the victim classifier. For example, \cite{chen2017zoo} proposed a method to estimate the gradient with score information and craft adversarial examples with the estimated gradient. \cite{ilyas2018black} introduced a method which leverages natural evolutionary strategy to estimate the gradient and generate adversarial examples. In general, score-based attacks can be divided into two sub-categories. 

{\bf Gradient-approximation based methods:} The first sub-category are gradient-approximation based methods.  As the name suggests, these methods first approximate the gradients or the signs of the gradients, then generate adversarial examples using the approximated information.

Zeroth Order Optimization Based Attack (ZOO) proposed by~\cite{chen2017zoo} uses a finite difference method to approximate the gradient of the loss w.r.t. the input. Then the C$\&$W attack is applied to generate an adversarial example. It uses the following formula to estimate the gradient:
\begin{align*}
    \frac{\partial f(\bx)}{\partial \bx_{(i)}}\approx \frac{f(\bx+h\be_i)-f(\bx-h\be_i)}{2h},
\end{align*}
where $h$ is a small constant and $\be_i$ is a standard basis vector with only the $i$th component as $1$, and $i$ ranges from $1$ to the dimension of input. The time used to estimate the gradients grows with the dimension. When the dimension of the input is large, the authors introduced several techniques to scale-up the estimation, making it possible for the method to craft adversarial examples in reasonable time for large DNNs trained on large datasets, such as ImageNet~\citep{deng2009imagenet}.

Many other methods are introduced to efficiently approximate the gradient and generate adversarial examples based on it, such as NES attack~\citep{ilyas2018black} and Bandits Attack~\citep{ilyas2018prior}. NES attack uses natural evolutionary strategies (hence the acronym NES) to estimate the gradient of loss w.r.t. the input then generates adversarial examples based on the estimated gradient:
\begin{align*}
    \frac{1}{\sigma n}\sum_{i=1}^n\BD_if(\bx +\sigma\BD_i),
\end{align*}
where $n$ is the number of searches to estimate gradient, $\BD_i$ is random direction sampled from $\mathcal{N}(0, I)$ and $\sigma$ is the standard deviation of the search step.
The authors also extended the method to a partial-information setting, where only part of the scores or top-$k$ sorted labels are given.



In fact, the most important information for adversarial attack is the sign of the gradient, which provides the direction of optimization. Many methods are proposed to estimate the sign of the gradient and use that information to craft adversarial examples; notable examples include SignSGD~\citep{liu2018signsgd} and Sign hunter~\citep{al2019there}.  To further improve the efficiency of gradient approximation or sign estimation, some researchers propose using a substitute model, which is a model trained with the same data as the victim model and performs similarly to the victim model.  Representative examples include Subspace attack~\citep{yan2019subspace} and Transfer-based prior~\citep{cheng2018query}. Also, \cite{wang2020spanning} shows how to exploit the data distribution to identify importance subspace for black-box attacks. These attack methods also belong to the gradient-approximation based category since their key element is to approximate gradient information efficiently.

{\bf Other methods:} The second sub-category of score-based methods consists of methods that do not approximate gradient related information to generate adversarial examples. For example, \cite{li2018nattack} proposed the Gaussian black-box adversarial attack (Nattack), which searches the adversarial examples by modeling the adversarial population with a Gaussian distribution. The intuition comes from the fact that one can find various adversarial examples for the same input using different attack methods, which suggests there is a population of adversarial examples.  Other methods that fall into this sub-category include the GenAttack~\citep{alzantot2018genattack},  the Simple Black-Box Attack~\citep{guo2019simple} and Square attack~\citep{andriushchenko2020square}.

\subsection{Decision-Based Attack}

In many practical situations, the attacker has access only to the predicted labels of the model, but not any gradient or score information. When only the predicted label $c(\bx)$ is available, both gradient-based and score-based methods do not work. 
\cite{papernot2017practical} introduced a transfer attack method which requires only the observations of the labels predicted by the model. The main idea is to train a substitute model which is similar to the target model and attack the substitute model instead. Boundary Attack was subsequently proposed by~\cite{brendel2017decision}, which searches for adversarial examples based on random-walk on the decision boundary. Many extensions of Boundary Attack have been proposed to improve the efficiency and performance of it~\citep{brunner2019guessing,chen2019boundary,chen2019hopskipjumpattack,guo2018low}.
{\color{black}There are also decision-based attack methods that are neither transfer-based nor random-walk based~\citep{ilyas2018black,cheng2019sign, yan2019subspace}. In general, the goal of decision-based attack is to generate adversarial examples with the predicted labels returned by the victim model. Decision-based attack methods can be divided into three sub-categories.}

{\bf Transfer based attacks:} The first category consists of transfer-based attack methods. Various researchers have observed that, if two DNNs are trained with similar data, even though the two models may have very different structures, adversarial examples generated on one model can be used to fool another one. Based on this observation, \cite{papernot2017practical} proposed to train a substitute model based on a small amount of training data and generate adversarial examples based on the substitute model. They showed that the adversarial examples generated on the substitute model can also fool the target classifier. The proposed method does not require too many samples to train the substitute model and can achieve relatively high success rate in untargeted task. However, the method does not work well in targeted task. \cite{liu2016delving} proposed an ensemble-based approach to generating transferable adversarial examples. This ensemble approach increases the proportion of target adversarial examples transferring with their target labels. 

\begin{figure}
    \centering
    \includegraphics[width=0.6\textwidth]{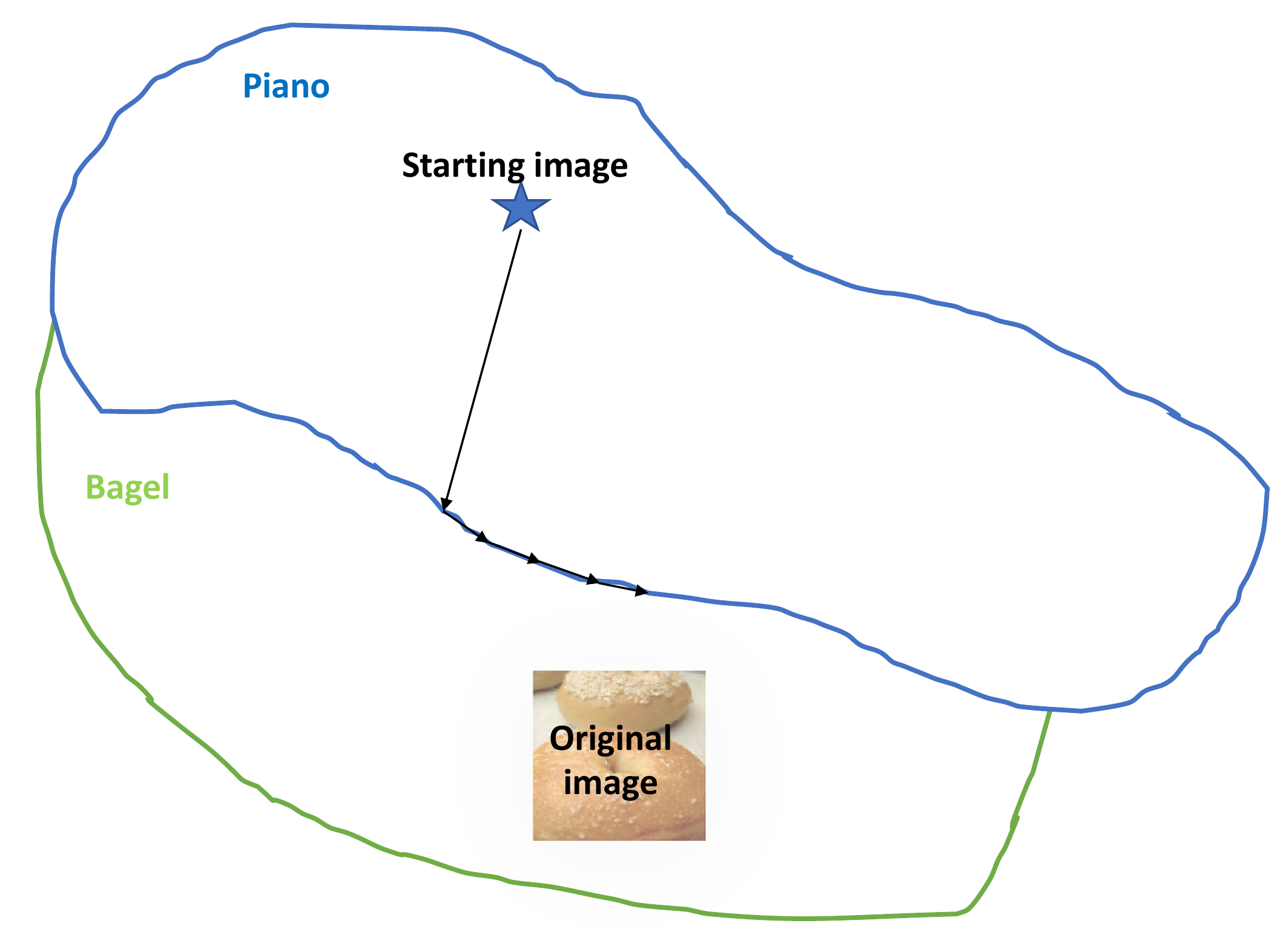}
    \caption{An illustration of Boundary Attack. In essence the Boundary Attack performs rejection sampling along the boundary between adversarial and non-adversarial images~\citep{brendel2017decision}.}
    \label{fig:boundary_attack}
\end{figure}

{\bf Random-walk based attacks:} The second category consists of methods that are based on random walk on the boundary. \cite{brendel2017decision} proposed Boundary Attack, a method that does not rely on the gradient of loss w.r.t. the input and performs well under both targeted and untargeted settings. {\color{black}The method starts with a sample categorized in the targeted class and seeks to minimize the perturbation while staying adversarial.} The process of boundary attack is shown in Figure~\ref{fig:boundary_attack} and details of the method are described in Algorithm~\ref{alg:battack}.

\alglanguage{pseudocode}
\begin{algorithm}[H]
\caption{Boundary Attack}
\label{alg:battack}
\begin{algorithmic}[1]
\State \textbf{Input:} decision model $c(\cdot)$, original input $\bx$ 
\State \textbf{Output:} $\bx^{adv}$
\State \textbf{Initialization:} $t=0$, $\tilde{\bx}^0\sim\mU(0,1)$ s.t. $c(\tilde{\bx}^0)=$target class
\While {$t<T$}
\State $t=t+1$
\State draw random perturbation from proposal distribution $\boldsymbol{\eta}^t\sim \mathcal{P}(\tilde{\bx}^{t-1})$
\If {$c(\tilde{\bx}^{t-1}+\boldsymbol{\eta}^t)=$target class}
\State set $\tilde{\bx}^t=\tilde{\bx}^{t-1}+\boldsymbol{\eta}^t$
\Else
\State set $\tilde{\bx}^t=\tilde{\bx}^{t-1}$
\EndIf
\EndWhile
\State \textbf{Return} $\tilde{\bx}^{T}$
\end{algorithmic}
Note that $c(\bx)$ is the label assigned to the input $\bx$ by the classifier, and $T$ is the maximum number of iterations. 
\end{algorithm}

Define input space as $\mathcal{D}$, the process of drawing $\boldsymbol{\eta}^t$ from the proposal distribution is as follows:
\begin{itemize}
    \item Draw $\boldsymbol{\eta}^t\sim\mathcal{N}(\boldsymbol{0},\boldsymbol{I})$.
    \item Rescale and clip $\boldsymbol{\eta}^t$ to make sure $\tilde{\bx}^{t-1}+\boldsymbol{\eta}^t\in \mathcal{D}$ and $\|\boldsymbol{\eta}^t\|_2=\delta\cdot d(\bx,\tilde{\bx}^{t-1})$, where $d(\cdot,\cdot)$ represents distance function between two samples and $\delta$ is a hyper-parameter that controls the scale of the perturbation.
    \item Orthogonal perturbation: project $\boldsymbol{\eta}^t$ onto a sphere around the original input $\bx$ such that $d(\bx,\tilde{\bx}^{t-1}+\boldsymbol{\eta}^t)=d(\bx,\tilde{\bx}^{t-1})$.
    \item Move towards the original input such that $\tilde{\bx}^{t-1}+\boldsymbol{\eta}^t\in\mathcal{D}$ and $d(\bx,\tilde{\bx}^{t-1})-d(\bx,\tilde{\bx}^{t-1}+\boldsymbol{\eta}^t)=\epsilon\cdot d(\bx,\tilde{\bx}^{t-1})$ both hold, where $\epsilon$ controls the step size of the movement.
\end{itemize}

Boundary Attack achieves performance comparable to state-of-the-art white-box attacks on DNNs trained for classification. \cite{chen2019boundary} introduced Boundary Attack$++$, using binary information at the decision boundary to estimate gradient direction. They showed that Boundary Attack$++$ requires significantly fewer model queries than Boundary Attack. \cite{guo2018low} proposed a decision-based black-box attack, which improves the query-efficiency of Boundary Attack by restricting the search for adversarial examples to a low frequency domain. There are many other methods belong to this category, such as 
Guessing Smart~\citep{brunner2019guessing}.

{\bf Optimization based attacks: }
Instead of conducting random walks, a process which does not have any convergence guarantees, more recently researchers have found that attacks in the decision-based setting can also be formulated as solving a zeroth order optimization problem~\citep{cheng2019improving}. They showed that both PGD and C\&W loss are ill-defined in the decision-based setting, and instead we need to redefine the problem as finding the best direction $\btheta$ of adversarial examples. Given $\bx_0$, a function $g(\cdot)$ that measures the distance between $\bx_0$ and the decision boundary along direction $\btheta$ can be defined as
\begin{equation}
    g(\btheta) = \arg\min_{\lambda > 0 }\big( f(\bx_0+\lambda \frac{\btheta}{\|\btheta\|})\neq y_0 \big). 
\end{equation}
The attack problem can then be posed as
\begin{equation}
    \btheta^* = \arg\min_{\btheta} g(\btheta), 
    \label{eq:opt_attack}
\end{equation}
and the adversarial example that is closest to $\bx_0$ is $\bx^* = \bx_0 + g(\btheta^*) \frac{\btheta}{\|\btheta\|}$. Although the gradient of $g(\btheta)$ cannot be directly computed, the function value of $g(\btheta)$ can be computed by binary search, thus standard zeroth order optimization solvers can be applied for solving \eqref{eq:opt_attack}. In OPT attack~\citep{cheng2018query}, Randomized Gradient-Free (RGF) method~\citep{nesterov2017random} is used to address the problem. 
Later, \cite{cheng2019sign} showed that the gradient sign of \eqref{eq:opt_attack} can be computed in a more query-efficient way, leading to an improved attack called Sign-OPT. On the other hand, \cite{chen2019hopskipjumpattack} proposed another optimization-based formulation, leading to yet another query-efficient algorithm called HotSkipJump Attack. 
Recently, \cite{chen2020rays} proposed RayS, which reformulates the continuous optimization problem in~\cite{cheng2018query} into a discrete one for $\ell_\infty$ norm attack. 

\section{Defense}
\label{sec:defense}

There has been extensive research on improving the robustness of DNNs for defending against adversarial examples. In general, methods that aim to increase model robustness fall into the following four main categories: (1) augmenting the training data with adversarial examples, (2) leveraging randomness to defend against adversarial attacks, (3) removing adversarial perturbations with projection, and (4) detecting the adversarial examples instead of classifying them correctly.

\subsection{Adversarial Training}

To improve the robustness of a classifier against adversarial examples, one natural idea is to train the model with adversarial examples. \cite{goodfellow2014explaining} proposed generating adversarial examples with FGSM attack and adding them back to the training data set to improve the robustness of the model. Later \cite{kurakin2016adversarial} suggested using a multi-step FGSM to further improve the adversarial robustness. 

\cite{madry2017towards} introduced a min-max formulation, which iteratively generates adversarial examples with PGD attack while training the model. \cite{athalye2018obfuscated} showed that {\bf Madry's adversarial training} can survive strong attacks while many other state-of-the-art defense methods cannot. Instead of minimizing the original loss function $L(\BT,\bx,y)$, 
{\color{black} the following min-max objective function is used:}
\begin{align*}
    &\min\limits_{\BT} \max\limits_{\BD\in S}L(\BT,\bx+\BD,y).
\end{align*}
The inner maximization problem aims to find an adversarial perturbation of a given data point $\bx$ within the perturbation set $S$ that achieves a high loss. Indeed, this is precisely the problem of attacking any given neural network. On the other hand, the goal of the outer minimization problem is to find the model parameters that minimize the ``adversarial loss'' given by the inner attack problem. Therefore, the model is trained with the adversarial example in the $\epsilon$-ball of the original sample. Recently, it has been shown that such min-max optimization is guaranteed to converge under certain assumptions~\citep{gao2019convergence}. 

Many defense methods based on Madry's adversarial training have been proposed to further improve its efficiency and performance. For example, \cite{zhang2019theoretically} proposed TRADES, a theoretically-driven upper bound minimization algorithm that achieved the top-1 rank in the NeurIPS 2018 defense competition. 


In other methods, researchers treated mis-classified examples and correctly classified examples as the same when performing adversarial training. \cite{ding2018max} noticed the importance of mis-classified examples and improved the performance of adversarial training by combining the usual cross-entropy loss with a margin maximization loss term applied to the correctly classified examples. \cite{wang2020improving} found that mis-classified examples have more impact on the final robustness than correctly classified examples and incorporate mis-classified examples in adversarial training as a regularizer.

Other than just adding adversarial examples into the training process, \cite{wang2019convergence} studied the convergence quality of adversarial examples found in the inner maximization and proposed a dynamic adversarial training method that changes adversarial strength in inner maximization according to convergence quality. The convergence of adversarial training has also been studied in many other works~\citep[e.g.,][]{gao2019convergence}. 

Adversarial training performs well against attacks but generating adversarial examples during training could be expensive. Consequently, various researchers have developed methods for reducing the computational overhead brought by adversarial training~\citep{shafahi2019adversarial,zhang2019you,wong2020fast}. Many extensions of adversarial training have also been proposed to improve the performance of adversarial training~\citep{carmon2019unlabeled,zhang2019theoretically,gowal2020uncovering,wu2020adversarial,pang2020boost,pang2021bag}. 

\subsection{Randomization}

Another group of effective defense methods leverages randomness to defend against adversarial examples. Adversarial perturbation can be viewed as noise, and various methods have been proposed to improve the robustness of DNNs by incorporating random components into the model. The randomness can be introduced to:
\begin{itemize}
\item the input; i.e., randomizing the input of a neural network to remove the potential adversarial perturbation~\citep[e.g.,][]{xie2017mitigating,cohen2019certified},
\item the hidden layer output; i.e., adding Gaussian noise to the input and hidden output~\citep[e.g.,][]{liu2017towards} and introducing pruning methods to randomize the network output~\citep[e.g.,][]{dhillon2018stochastic}, or 
\item the parameters of the classifier; i.e., leveraging a Bayesian component to add randomness to the weights of the model~\citep[e.g.,][]{liu2018advbnn}.
\end{itemize}
The robustness of neural networks against random noise has been analyzed both theoretically~\citep{fawzi2016robustness,franceschi2018robustness} and empirically~\citep{dodge2017study}. 

\cite{xie2017mitigating} introduced a simple preprocessing method to randomize the input of neural networks, hoping to remove the potential adversarial perturbation. During the testing phase, the input is randomly resized to several sizes, then around each of the resized inputs, zeros are randomly padded. The authors demonstrated that this simple method can be applied to large-scale datasets, such as Imagenet.  Similarly, \cite{zantedeschi2017efficient} showed that, by using a modified ReLU activation layer (called BReLU) and adding noise to the origin input to augment the training data, the learned model will gain some stability to adversarial examples. However, other researchers have found that this defense method is not robust against strong white-box attack, such as PGD and C$\&$W~\citep{carlini2017towards}.

Instead of adding random components to the input space, \cite{liu2017towards} proposed a ``noise layer'' to introduce randomness to both the input and the hidden layer output. In this ``noise layer'', randomly generated Gaussian noise is added to the input:
\[ \bx\leftarrow\bx+\boldsymbol{\varepsilon}, \quad \boldsymbol{\varepsilon}\sim\mathcal{N}(0,\sigma^2I),
\]
where $\sigma$ is a hyper-parameter. Larger values of $\sigma$ lead to better robustness but worse prediction accuracy, while smaller values of $\sigma$ result in better prediction accuracy but deteriorated robustness. The noise layer is applied in both training and testing phases, so the prediction accuracy will not be largely affected. The authors showed that Random Self-Ensemble (RSE) is equivalent to an ensemble possessing an infinite number of noisy models without any additional memory overhead. However, the experiments demonstrated that, although RSE can increase the robustness of DNNs, it also sacrifices a non-negligible amount of accuracy. Later, \cite{liu2018advbnn} introduced a new min-max formulation to combine adversarial training with Bayesian Neural Networks (BNNs). All weights in a Bayesian neural network are represented by probability distributions over possible values, rather than having a single fixed value. The proposed framework, called Adv-BNN, combines adversarial training with randomness, is shown to have significant improvement over previous approaches including RSE and Madry's adversarial training.

Although these methods demonstrate improvement of robustness against adversarial examples, they typically lack theoretical guarantee for their performances. \cite{lecuyer2019certified, li2018certified,cohen2019certified} introduced a method called {\bf Randomized Smoothing} and proved a tight robustness guarantee in the $\ell_2$ norm. The details of Randomized Smoothing are summarized below:
\begin{itemize}
    \item Training stage:
    \begin{enumerate}
        \item Train the base classifier with Gaussian data augmentation with variance $\sigma^2$.  That is, the base classifier is trained on the natural examples and noisy examples, where the noisy examples are generated by adding Gaussian noise to natural examples.
        \item The trained base classifier is denoted as $f$.
    \end{enumerate}
    \item Testing stage: for a given input testing sample $\bx$:
    \begin{enumerate}
        \item Generate $n$ noise corrupted samples by adding Gaussian noise with variance $\sigma^2$ to the input $\bx$.
        \item Feed the $n$ noise corrupted samples into the trained base classifier $f$ and obtain the top-2 frequent predicted classes from the predictions. The top-2 frequent classes are denoted as $\hat{c}_A, \hat{c}_B$ and the corresponding frequencies are denoted as $n_A, n_B$.
        \item If BINOMPVALUE$(n_A,n_A+n_B,0.5)\le \alpha$, the final predicted class is $\hat{c}_A$, otherwise abstain. The function BINOMPVALUE$(n_A,n_A+n_B,p)$ returns the $p$-value of the two-sided hypothesis test that $n_A\sim$Binomial$(n_A+n_B,p)$ and $\alpha$ is the abstention threshold generated based on~\cite{hung2019rank}.
    \end{enumerate}
\end{itemize}
They also provided a process to generate certificated radius around the input. Theoretical analysis was provided to show that the classifier is robust around the input within the found $\ell_2$ radius.

\subsection{Projection}

The development of generative models, such as auto-encoders and generative adversarial networks, gives rise to another line of research, which removes adversarial noise by fitting generative models on the training data~\citep{jalal2017robust,meng2017magnet,li2018optimal,samangouei2018defense}. 
These defense mechanisms employ the power of generative models to combat the threat from adversarial examples. The input to the classifier is first fed into the generative model and then classified. Since the generative model is trained on natural examples, adversarial examples will be projected to the manifold learned by the generative model. Furthermore, ``projecting'' the adversarial examples onto the range of the generative model can have the desirable effect of reducing the adversarial perturbation.

\cite{meng2017magnet} introduced MagNet as a robust framework against adversarial examples. It trains multiple auto-encoders to move adversarial examples closer to the manifold of natural examples. An auto-encoder is a type of neural network that consists of two major parts: the encoder that maps the input to a low-dimensional space and the decoder that recovers the input from the low-dimensional embedding. The auto-encoder is usually trained on the reconstruction loss w.r.t. the input. Therefore, the high-dimensional data is summarized by the low-dimensional embedding through the training process. In MagNet, one auto-encoder is chosen at random at testing time to filter the input samples, and thus adversarial perturbation could potentially be removed through this encoding and decoding process.

Instead of using auto-encoders, \cite{samangouei2018defense} proposed to train a Generative Adversarial Networks (GAN)~\citep{goodfellow2014explaining} to fit the training data distribution and help remove adversarial noise. It first trains a GAN to model the distribution of the training data. Then, at inference time, it finds an output of the GAN that is close to the input, and feeds that output into the classifier. This process ``projects'' input samples onto the range of the GAN's generator, which can potentially help remove the effect of adversarial perturbations. Here the GAN is proposed to estimate the generative model through an adversarial process, in which two major parts are simultaneously trained: a generative model $\bG$ that captures the data distribution, and a discriminative model $\bD$ that estimates the probability that a sample came from the training data rather than $\bG$. The objective of training a GAN is:
\begin{align*}
    \min\limits_\bG\max\limits_\bD\left\{\E_{\bx\sim p_{data}(\bx)}\left[\log\bD(\bx)\right]+\E_{\bz\sim p_\bz(\bz)}\left[\log(1-\bD(\bG(\bz)))\right]\right\},
\end{align*}
where $p_{data}(\bx)$ represents the data distribution and $p_\bz(\bz)$ represents a prior distribution that is predefined. Therefore, $\bD$ and $\bG$ together play a two-player min-max game to recover the data distribution.

Defense-GAN is similar to MagNet as both of these methods aim to filter out the adversarial noise by using a generative model. \cite{samangouei2018defense} showed that Defense-GAN is more robust than MagNet on several benchmark datasets. There are some other works also follow similar patterns~\citep{jalal2017robust,song2017pixeldefend}.

\begin{figure}
    \centering
    \includegraphics[width=0.8\textwidth]{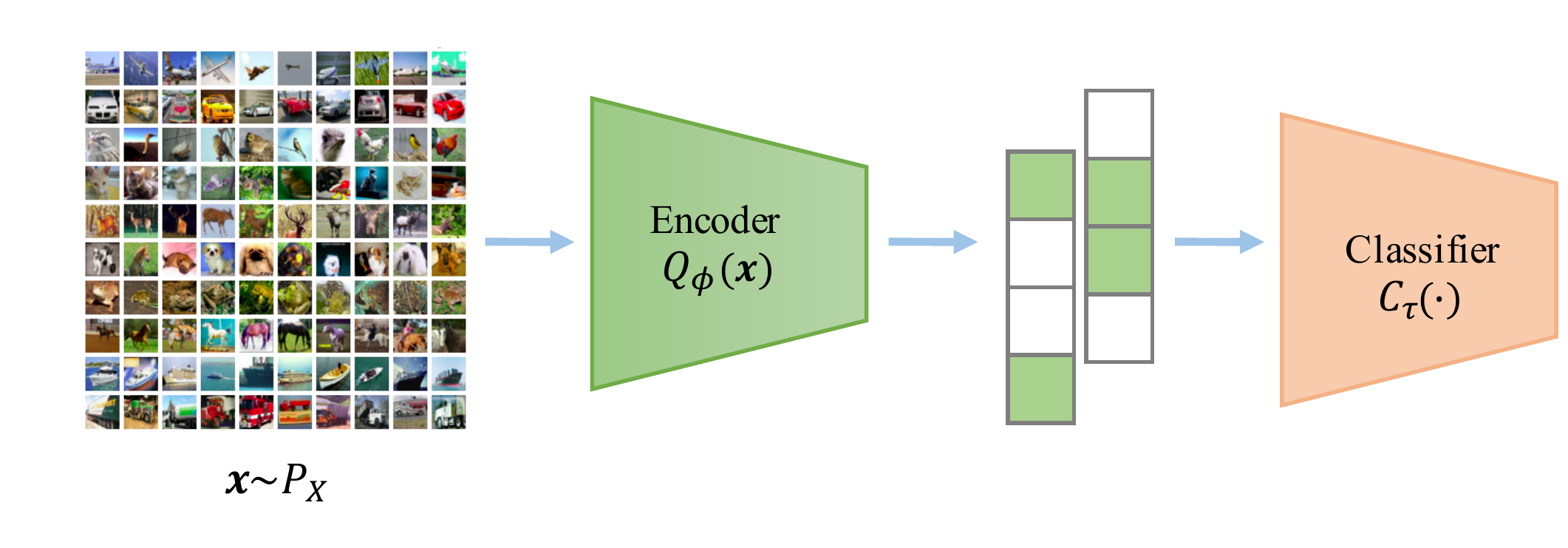}
    \caption{A deep classifier.}
    \label{fig:d_cla}
\end{figure}

All of the above methods train a separate generative model to perform the adversarial noise filtering. \cite{li2018optimal} proposed a different defense framework, termed {\bf ER-Classifier}, which combines the process of filtering and classification in one framework. In fact, any deep classifier can be viewed as a combination of these two parts as illustrated in Figure~\ref{fig:d_cla}: an encoder part to extract useful features from the input data and a classifier part to perform classification based on the extracted features. Both the encoder and the classifier are neural networks.

\begin{figure}
    \centering
    \includegraphics[width=0.8\textwidth]{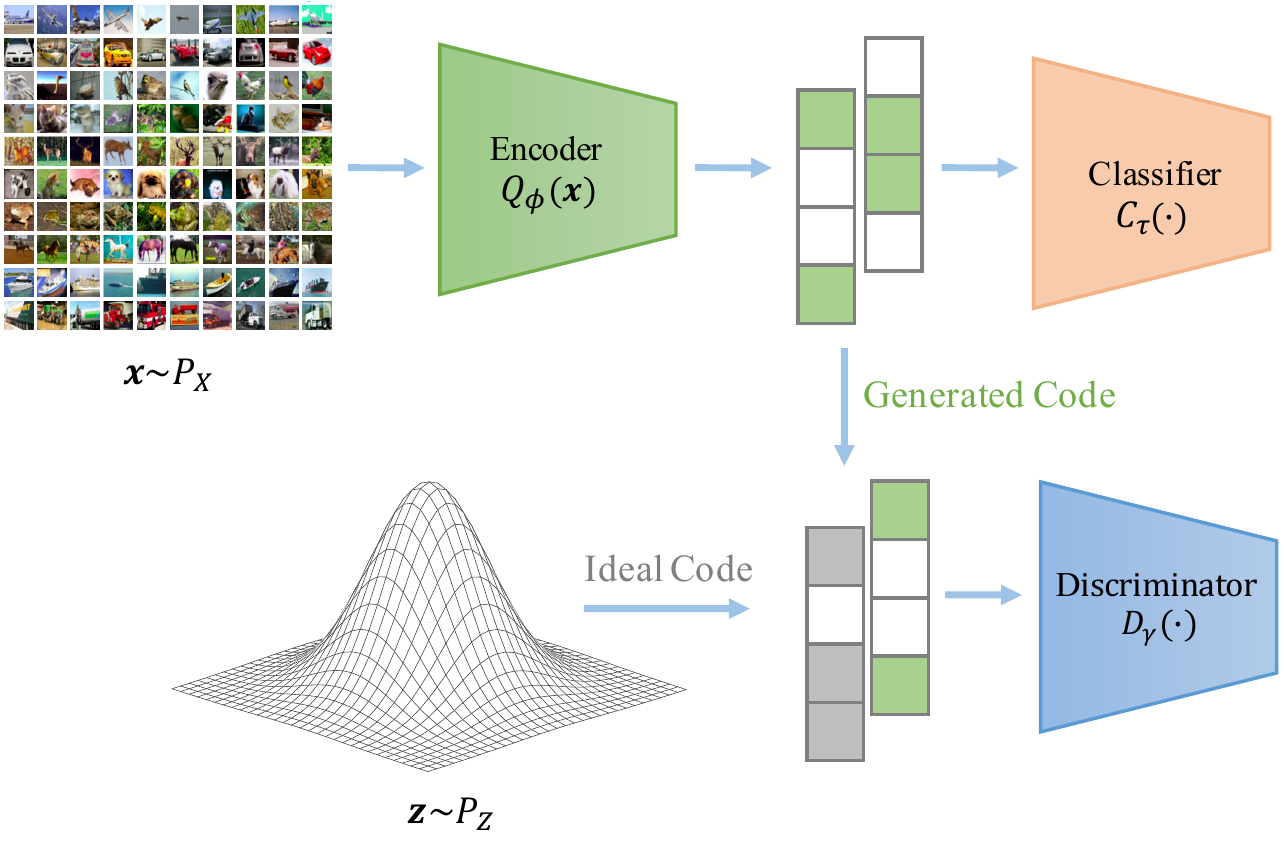}
    \caption{Overview of ER-Classifier framework.}
    \label{fig:er}
\end{figure}

ER-Classifier is similar to a regular deep classifier, which first projects the input to a low-dimensional space with an encoder $\bG$, then performs classification based on the low-dimensional embedding with an classifier $\bC$. The novelty is that at the training stage, the low-dimensional embedding of ER-Classifier is stabilized with a discriminator $\bD$ by minimizing the dispersion between the distribution of the embedding and the distribution of a selected prior. The goal of the discriminator is to separate the true code sampled from a prior and the ``fake'' code produced by the encoder, while the encoder will try to produce generated code that is similar to the true one.  The result of this competition is that the distribution of the embedding space will be pushed towards the prior distribution.  Therefore it is expected that this regularization process can help remove the effects of any adversarial distortion and push the adversarial examples back to the natural data manifold. Another difference is that the embedding space dimension is much smaller for the ER-classifier, when compared with a general deep classifier, making it easier for the training process to converge. Details of the framework are illustrated in Figure~\ref{fig:er}.
\cite{li2018optimal} showed that ER-Classifier achieves state-of-the-art performance against strong adversarial attack methods on several benchmark datasets.

\subsection{Detection}

Methods in this fourth category aim to detect the existence of adversarial examples, rather than trying to classify them into the correct classes. The main assumption behind these methods is that the adversarial examples came from a different distribution when comparing to those from the natural data. In other words, adversarial examples do not lie on the data manifold, and DNNs classify correctly only the samples near the manifold of training data~\citep{tanay2016boundary}. Therefore adversarial examples may be detected with carefully designed detectors. 

A straightforward way towards adversarial example detection is to build a simple binary classifier separating the adversarial examples apart from the clean data~\citep{gong2017adversarial}. The advantage is that it serves as a preprocessing step without imposing any assumptions on the model it protects. However, this method suffers from a generalization limitation. In a similar method, a small ``detector'' subnetwork is trained as an augmentation of DNNs on the binary classification task to distinguish true data from adversarial perturbations~\citep{metzen2017detecting}, where the inputs (to this detector) are taken from intermediate layers of the classification network. Alternatively, a cascade classifier may be built on the outputs from the convolutional layers~\citep{li2017adversarial} to efficiently detect adversarial examples. \cite{zheng2018robust} proposed modeling the output of a classifier with Gaussian mixture models and performing hypothesis testing to detect adversarial examples. \cite{roth2019odds} showed that adversarial examples exist in cone-like regions in very specific directions from their corresponding natural inputs and proposed a new test statistic to detect adversarial examples with the findings. In a recent work, \cite{yang2020ml} studied the feature attributions of adversarial examples and proposed a detection method based on feature attribution scores.

Instead of building a detector based on input or output of the network, another line of research proposes to study the characteristics of adversarial examples and build detectors based on such features~\citep{feinman2017detecting,grosse2017statistical,lee2018simple,ma2018characterizing,raghuram2020detecting}. In order to detect adversarial examples that lie far from the data manifold, \cite{feinman2017detecting} proposed {\bf KD-Detection} to perform density estimation on the training data in the feature space of the last hidden layer to help detect adversarial examples. 

Kernel density estimation is used to measure how far a sample is from the provided data manifold.  Suppose that $\bx_1,...,\bx_n$ are training samples drawn from an unknown probability density $p_X(\bx)$.  Given $\bx$, we can use the following function to estimate the density score at $\bx$:
\begin{align*}
    \hat{p}_X(\bx)=\frac{1}{n}\sum_{i=1}^n K_\sigma(\bx,\bx_i),
\end{align*}
where $K_\sigma(\cdot, \cdot)$ is a kernel function. In the KD-Detection framework, one kernel density model is fitted for each class. Therefore, if $\bx$ is predicted with label $y$, the samples $\{\bx_i\}_{i=1}^n$ used to do the estimation are training samples from class $y$. After the kernel density is fitted, both natural examples and adversarial examples are fed into the model and density function to generate density scores. A logistic regression model is then fitted on the kernel density scores to detect whether the input is adversarial or not.

Many recent works fall into this category, studying the properties of adversarial examples and detecting them with hidden layer features. \cite{ma2018characterizing} observed Local Intrinsic Dimension (LID) of hidden-layer outputs differ between the original and adversarial examples, and leveraged these characteristics to detect adversarial examples. In another work, \cite{lee2018simple} generated the class conditional Gaussian distributions with respect to hidden layer output of the DNN under Gaussian discriminant analysis, which result in a confidence score based on the Mahalanobis distance (MAHA), followed by a logistic regression model on the confidence scores to detect adversarial examples. Recently, a joint statistical test pooling information from multiple layers is proposed in~\cite{raghuram2020detecting} to detect adversarial examples. 

\section{Numerical Experiments}
\label{sec:exp}
In this section, we show how the attack methods fool various classifiers, including classical models like logistic regression and more recent models like DNNs. In the first part, the results of attacking a logistic regression model with projected gradient descent attack are shown. In the second part, we select representative attack methods from each category (gradient-based, score-based, and decision-based) and compare their performances. In the last part, we select one defense method from each category (adversarial training, randomization, projection, and detection) and compare their performances. The following benchmark datasets are used in our experiments:
\begin{itemize}
    \item MNIST~\citep{lecun1998mnist}: a handwritten digit dataset that consists of $60,000$ training images and $10,000$ testing images. These are $28\times 28$ black and white images in ten different classes.
    \item CIFAR10~\citep{krizhevsky2009learning}: a natural image dataset that contains $50,000$ training images and $10,000$ testing images with ten different classes. These are low resolution $32\times 32$ color images.
\end{itemize}
The input image is stored in RGB format with three channels, where the pixel values of each channel range from $0$ to $256$. We use the $\ell_\infty$ and $\ell_2$ distortion metrics to measure similarity between images and report the $\ell_\infty$ distance in the normalized $[0,1]$ space (e.g., a distortion of $0.031$ corresponds to $8/256$), and the $\ell_2$ distance as the total root-mean-square distortion normalized by the total number of pixels. Computing codes for carrying out the following numerical experiments can be downloaded at \url{https://github.com/reviewadvexample/revew_adv_defense}. 

\subsection{Attacking Logistic Regression Model}
We train a logistic regression model to perform binary classification on two handwritten digits, zero and one. The model is trained on all the images with label $0$ or $1$ in the training dataset and tested on all the images with label $0$ or $1$ in the testing dataset. Therefore, the training and testing sets are subsets of the training and testing sets of MNIST. There are $5,923$ handwritten-zero images and $6,742$ handwritten-one images in the training set. For testing set, there are $980$ handwritten-zero images and $1,135$ handwritten-one images.

The loss function of the logistic regression model used is:
\begin{align}
    f(\bw)=\sum_{i=1}^n\log(1+\exp(-y_i\bw^T\bx_i)),
\end{align}
where $\bw\in\R^{784}$ is the weight, $\bx_1, \bx_2, ..., \bx_n\in\R^{784}$ are input images vectorized from the original $28\times 28$ pixel space dimension, and $y_1, y_2, ..., y_n\in\{-1,1\}$ are labels with $-1$ corresponding to handwritten-zero and $1$ corresponding to handwritten-one.

The projected gradient descent (PGD) method attacks the model using gradient information. The gradient of the loss w.r.t. a certain input image $\bx_i$ is:
\begin{align*}
    \nabla f(\bx_i) = -\frac{y_i}{1+\exp(y_i\bw^T\bx_i)}\bw. 
\end{align*}
PGD attacks the model by repeating the following updating equation:
\begin{align*}
    \bx^{t+1}=\Pi_\epsilon\left\{\bx^t+\alpha\cdot\text{sign}\left(\nabla f(\bx_i)\right),\bx_i\right\}.
\end{align*}
In our experiment, we set $\alpha=1$, $\epsilon=8$, which reflects $0.031$ in $[0,1]$ scale since $\frac{8}{256}\approx 0.031$, and run the updates for $40$ steps.

\begin{figure}
    \centering
    \includegraphics[width=0.4\textwidth]{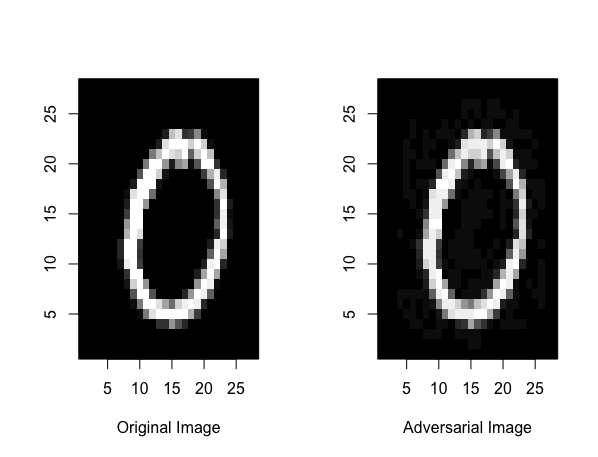}
    \includegraphics[width=0.4\textwidth]{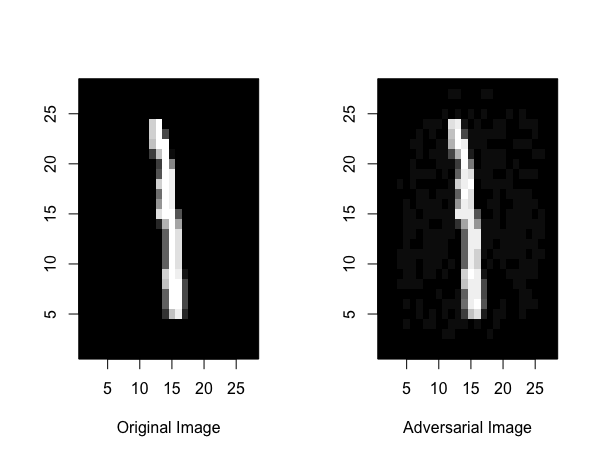}
    \includegraphics[width=0.4\textwidth]{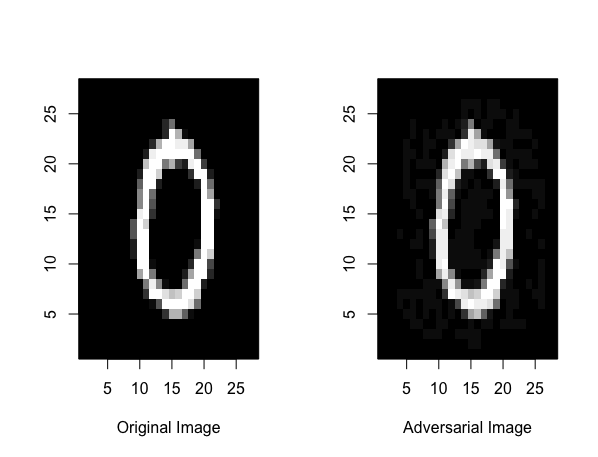}
    \includegraphics[width=0.4\textwidth]{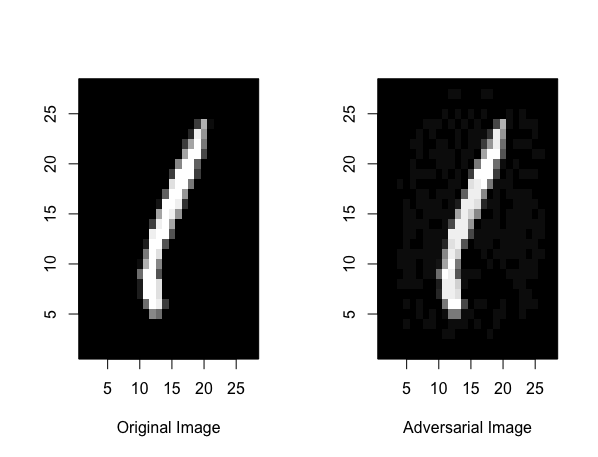}
    \caption{Four examples of attack results on a logistic regression model. In each pair of images, the left shows the original handwritten image and the right shows the adversarial image, which is classified incorrectly by the logistic regression model.}
    \label{fig:visual_lr_pgd}
\end{figure}

The prediction accuracy under no attack is $99.95\%$. Under PGD attack, the prediction accuracy drops to $72.96\%$. In Figure~\ref{fig:visual_lr_pgd}, we visualize four attack results, where in each pair the image on the left side is the original image and the image on the right side is the adversarial example based on the left one. We observe that the adversarial examples look remarkably similar to the original images but the predictions on these examples are altered by some invisible adversarial noise.

\subsection{Comparison of Different Attack Methods}
In this subsection we compare the performances of several attack methods including gradient-based, score-based and decision-based. The gradient-based attacks are compared on all samples from the test sets of the two benchmark datasets described above. Due to the slow speed of generating adversarial examples with score-based and decision-based approaches, the corresponding methods are evaluated on $300$ random samples from each of the test sets.

For the dataset MNIST, we train a network consisting of two convolutional layers with, respectively, 20 and 50 filters with kernel size 5, each followed by $2 \times 2$ max-pooling, and a fully connected layer of size 500. For CIFAR10, we use the standard VGG-16 structure~\citep{simonyan2014very} with batch normalization. 

The attack methods tested are:
\begin{itemize}
    \item Gradient-based: FGSM~\citep{goodfellow2014explaining} and PGD~\citep{madry2017towards}.
    \item Score-based: ZOO~\citep{chen2017zoo} and Square attack~\citep{andriushchenko2020square}.
    \item Decision-based: Boundary attack~\citep{brendel2017decision}, OPT attack~\citep{cheng2018query} and Sign-OPT~\citep{cheng2019sign}.
\end{itemize}

\paragraph{Results on gradient-based attack.} The results of attacking the two classification models on MNIST and CIFAR10 with gradient-based methods are shown in Table~\ref{tab:g_attack}. The strength of attack is represented by $\epsilon$, which controls the maximum distortion level. The larger the value of $\epsilon$, the stronger the attack. It is easy to see that PGD is stronger than FGSM since PGD runs the optimization process with equation~\eqref{eq:pgd} multiple times to search for adversarial examples, while FGSM runs the optimization for only one step.

\begin{table}[htbp]
    \centering
    \begin{tabular}{c|c|c|c|c|c|c|c|c}    
    \toprule
        \multirow{2}{*}{Method} &\multicolumn{4}{c|}{MNIST} &  \multicolumn{4}{c}{CIFAR10} \\
        \cline{2-9}
        & Clean Acc & $\epsilon=0.1$ & $\epsilon=0.2$ & $\epsilon=0.3$  & Clean Acc  & $\epsilon=0.01$ & $\epsilon=0.02$ & $\epsilon=0.03$ \\    
        \hline
        FGSM & 98.58\% & 75.98\% & 17.66\% & 2.63\% & 93.34\% & 56.46\% & 50.85\% & 47.61\%\\
        \hline
        PGD & 98.58\% & 69.42\%& 6.47\% & 0.02\% & 93.34\% & 5.71\% & 0.12\% & 0.01\%  \\
    \bottomrule
    \end{tabular}
    \caption{Accuracy on MNIST and CIFAR10 under gradient-based attack.}
    \label{tab:g_attack}
\end{table}

\paragraph{Results on score-based attack.} The results of score-based attacks on MNIST and CIFAR10 are shown in Table~\ref{tab:s_attack}. For ZOO~\citep{chen2017zoo}, the maximum number of iterations (Max.iter) limits the number of searches used to perform gradient estimation. The larger the value, the better the approximation. Therefore, a large quantity of maximum number of iterations usually results in better performance. Similarly for Square attack~\citep{andriushchenko2020square}, the maximum number of iterations controls the number of random-walk steps to search for adversarial examples. Therefore, a large value usually leads to better performance. The authors of Square attack studied the property of adversarial distortion and leveraged it to perform the attack. The method achieves better performances with the same maximum number of iterations when compared to the ZOO.

\begin{table}[htbp]
    \centering
    \begin{tabular}{c|c|c|c|c|c}    
    \toprule
        & Method & Clean Acc & Max.iter$=100$ & Max.iter$=200$ & Max.iter$=500$  \\    \hline
        \multirow{2}{*}{MNIST} & ZOO & 99.33\% & 99.33\% & 99.33\% & 99.00\% \\
        \cline{2-6}
        & Square & 99.33\% & 73.33\%& 34.00\% & 16.67\% \\ \hline\hline
        \multirow{2}{*}{CIFAR10} & ZOO & 93.67\% & 93.67\% & 90.33\% & 76.67\% \\
        \cline{2-6}
        & Square & 93.67\% & 84.67\%& 70.67\% & 54.33\% \\        
    \bottomrule
    \end{tabular}
    \caption{Accuracy on MNIST and CIFAR10 under score-based attack.}
    \label{tab:s_attack}
\end{table}

\begin{figure}[ht]
\centering
\subfloat[Decision Attacks on MNIST]{\label{fig:mdleft}{\includegraphics[height=0.28\textheight]{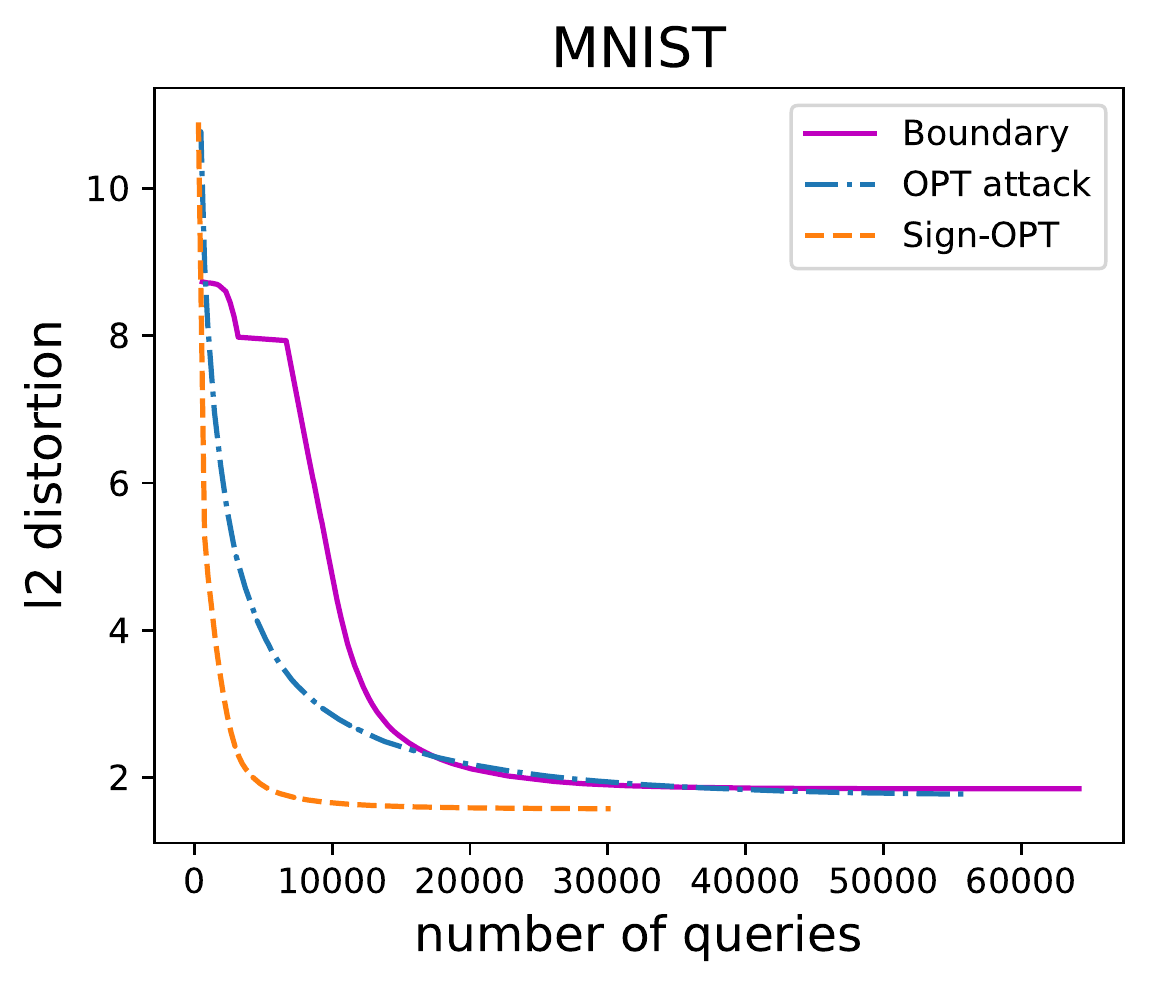}}}
\subfloat[Decision Attacks on CIFAR10]{\label{fig:mdright}{\includegraphics[height=0.28\textheight]{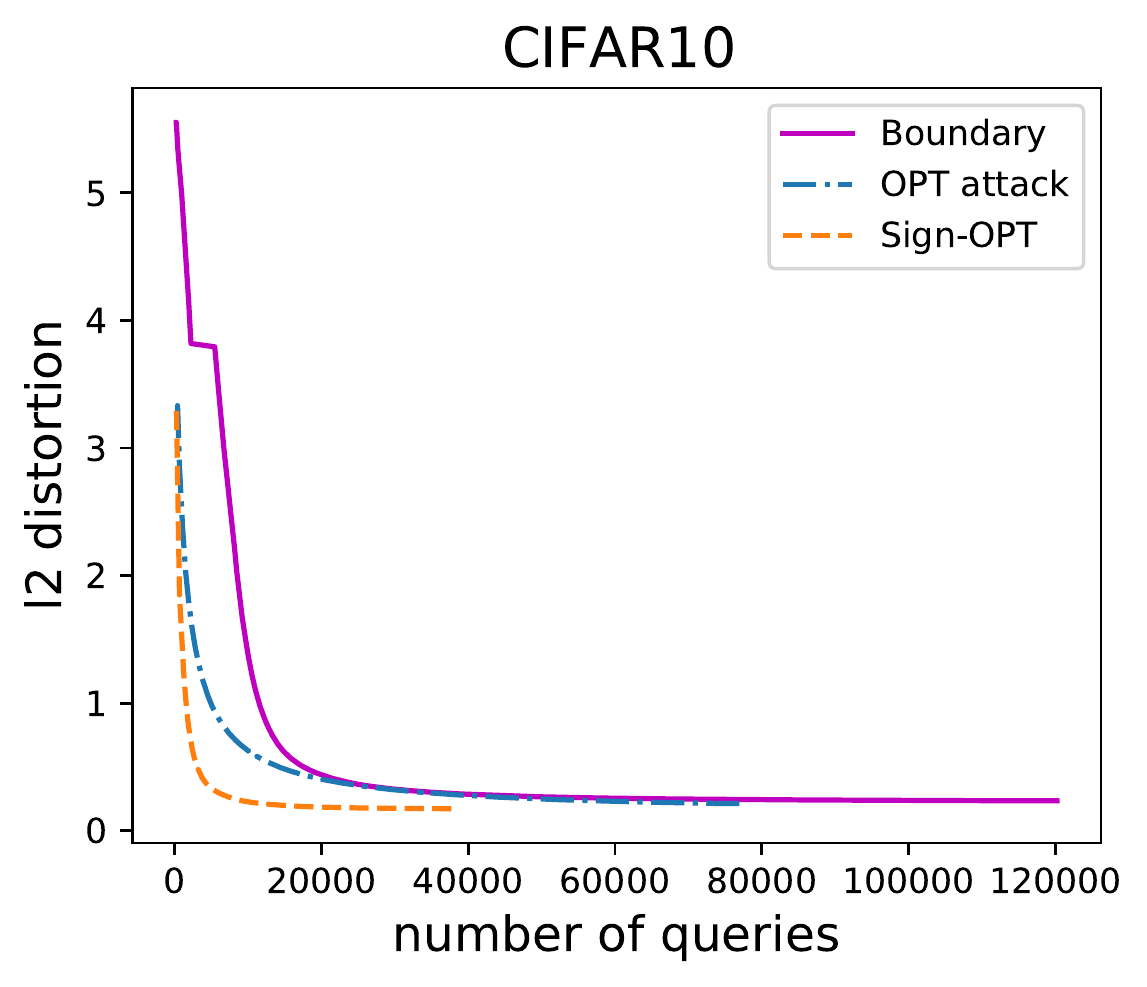}}}
\caption{Performances of decision-based attacks. Number of queries represents the number of times victim model is called.}
\label{fig:d_attack}
\end{figure}

\paragraph{Results on decision-based attack.} The results of decision-based experiments, in which only the predicted labels are available to the attacks, are shown in Figure~\ref{fig:d_attack}. Under this setting, only limited information is available, making it harder to craft adversarial examples. In Figure~\ref{fig:d_attack}, we compare Boundary~\citep{brendel2017decision}, OPT attack~\citep{cheng2018query} and Sign-OPT~\citep{cheng2019sign} by visualizing the relationship between the number of queries required versus distortion.  Boundary attack is an example of random-walk based attacks and the other two are examples of optimization based attacks.  The horizontal axis of each figure represents number of times the victim model is called to generate adversarial example. The vertical axis represents distortion level; i.e., the $\ell_2$ norm Euclidean distance between adversarial example and the original input. We observe that, under the same distortion level, Sign-OPT and OPT attacks require much fewer number of searches to craft adversarial examples than Boundary attack, and Sign-OPT is more efficient than OPT attack. 

\subsection{Comparison of Different Defense Methods}
We first compare defense methods from the first three categories (adversarial training, randomization and projection) and delay the last category (detection) towards the end of this subsection.  It is because detection methods are based on a different philosophy when comparing with the first three categories: they aim to detect the adversarial examples rather than to classify them correctly.  In other words, they are not directly comparable with other defense methods.  As before, we compare these methods with the same two benchmark datasets.  

The following methods from the first three categories are selected for comparison:
\begin{itemize}
    \item No Defense: Model without any defense method.
    \item Adversarial training: Madry's Adv~\citep{madry2017towards} and TRADES~\citep{zhang2019theoretically}.
    \item Randomization: RSE~\citep{liu2017towards} and Ran.Smooth~\citep{cohen2019certified}.
    \item Projection: ER-CLA~\citep{li2018optimal}.
\end{itemize}

\begin{figure}
    \centering
    \includegraphics[width=0.45\textwidth]{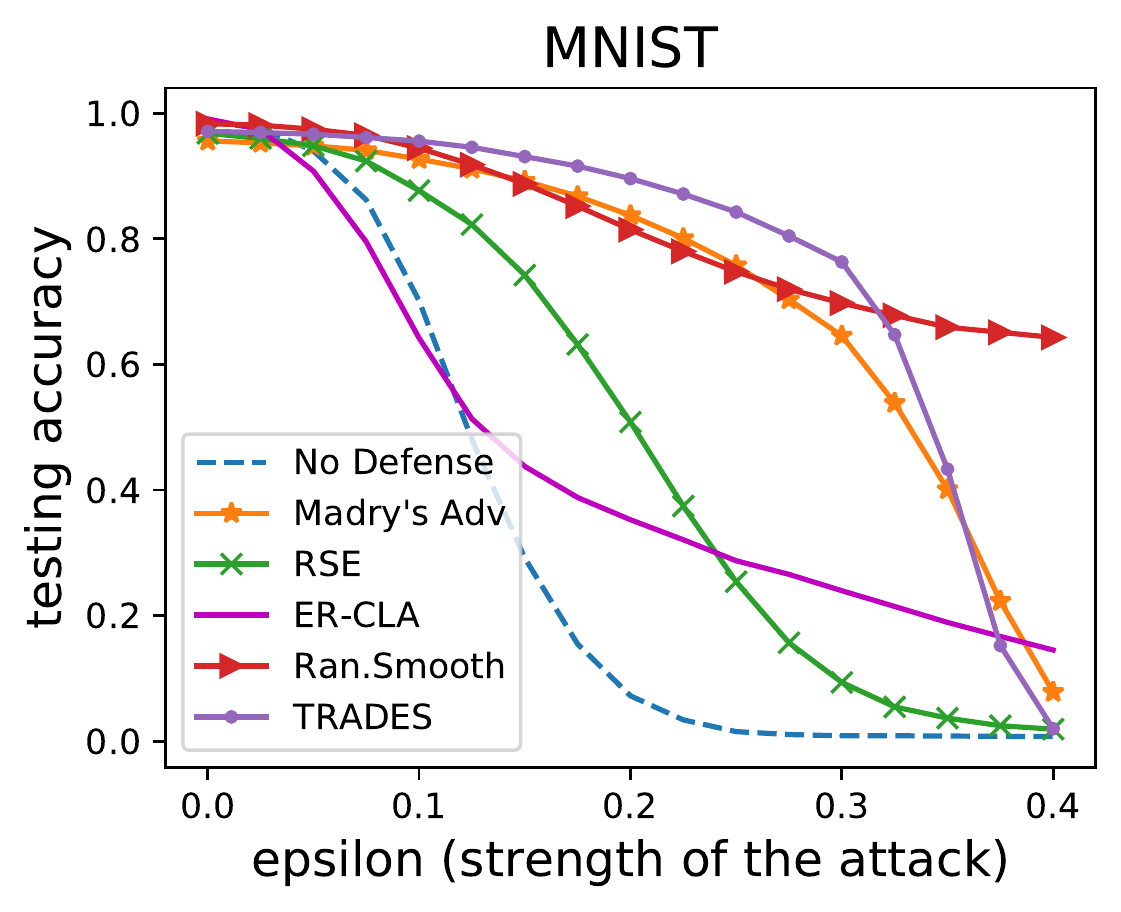}
    \includegraphics[width=0.45\textwidth]{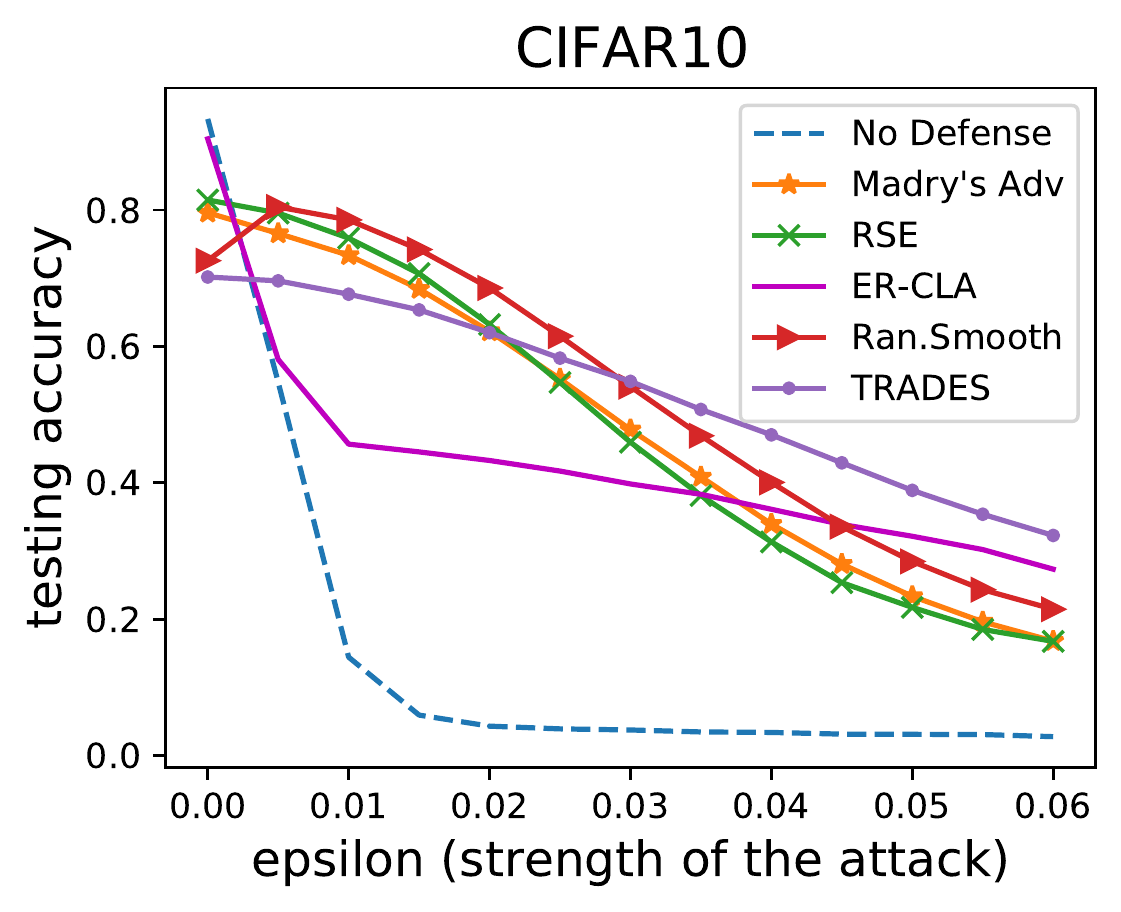}
    \caption{Comparison of Different Defense Methods on MNIST and CIFAR10}
    \label{fig:df_com}
\end{figure}

The comparison results of the first three categories of defense methods are shown in Figure~\ref{fig:df_com}. In the plot, the horizontal axis represents $\epsilon$, the strength of the PGD attack. The larger the $\epsilon$, the stronger the attack. The vertical axis represents the testing accuracy under the PGD attack. For all models, the accuracy drops as the attack strength increases except for Ran.Smooth on MNIST when the attack strength increases from $0$ to $0.05$. We observe that TRADES and Ran.Smooth perform better on MNIST and CIFAR10. On CIFAR10, RSE and Madry's adversarial training perform similarly, while TRADES performs best when attack is strong. Another point to notice is that, when there is no attack, all the other methods perform worse than ER-Classifier.  In fact, ER-Classifier and adversarial training methods can be combined together to further improve the results. In~\cite{li2018optimal}, the authors showed that combing ER-Classifier with adversarial training methods outperform both Madry's adversarial training and TRADES. All five defense methods perform much better than the model without any defense method.

Now we compare defense methods from the detection category.  The methods are:
\begin{itemize}
    \item KD: Kernel Density detection~\citep{feinman2017detecting}.
    \item LID: Local Ontrinsic Dimensionality detection~\citep{ma2018characterizing}. \item ODD: Odds are odd detection~\citep{roth2019odds}.
    \item ReBeL: Reading Between the Layers~\citep{raghuram2020detecting}. 
\end{itemize}

\begin{table}
    \centering
\resizebox{0.98\textwidth}{!}{
    \begin{tabular}{c|c|c|c|c|c||c|c|c|c||c|c|c|c}
    \toprule
    \multirow{2}{*}{Data}    & \multirow{2}{*}{Metric}  &   \multicolumn{4}{c||}{C$\&$W} & \multicolumn{4}{c||}{FGSM} & \multicolumn{4}{c}{PGD}    \\ \cline{3-14}
              & &  KD   &  LID  &  ODD & ReBeL &  KD   &  LID  &  ODD & ReBeL &  KD   &  LID  &  ODD & ReBeL \\ \hline \hline
    \multirow{4}{*}{CIFAR10} & AUC   & 0.945  & 0.947 &  0.955 &  {\bf 0.968} & 0.873 & 0.957  &  0.968 & {\bf 0.990} & 0.791 & 0.777 &  {\bf 0.963} & 0.962  \\ \cline{2-14}
               & TPR(FPR@0.01) & 0.068 & 0.220 & {\bf 0.591} & 0.309 & 0.136 & 0.385& 0.224 & {\bf 0.698} & 0.018 & 0.093 & 0.059 & {\bf 0.191}\\ \cline{2-14}
               & TPR(FPR@0.05) & 0.464 & 0.668 & {\bf 0.839} & 0.726 & 0.401 & 0.753 & 0.709 & {\bf 0.974} & 0.148 & 0.317 & {\bf 0.819} & 0.789 \\ \cline{2-14}
               & TPR(FPR@0.10) & 0.911 & 0.856 & 0.901 & {\bf 0.954} & 0.572 & 0.875 & {\bf 1.000} & {\bf 1.000} & 0.285 & 0.448 & {\bf 0.999} & {\bf 0.999} \\ \hline \hline
    \multirow{4}{*}{MNIST}   & AUC     & 0.932  & 0.785 &  0.968 & {\bf 0.980} & 0.933 & 0.888 &  0.952 & {\bf 0.992} & 0.801 & 0.861 &  0.967 & {\bf 0.975} \\ \cline{2-14}
               & TPR(FPR@0.01) & 0.196 & 0.079 & 0.212 & {\bf 0.630} & 0.421 & 0.152 & {\bf 0.898} & 0.885 & 0.062 & 0.170 & {\bf 0.607} & 0.382\\ \cline{2-14}
               & TPR(FPR@0.05) & 0.616 & 0.263 & {\bf 0.911} & 0.900 & 0.692 & 0.503 & 0.908 & {\bf 0.990} & 0.275 & 0.396 & {\bf 0.934} & 0.851 \\ \cline{2-14}
               & TPR(FPR@0.10) & 0.818 & 0.397 & {\bf 1.000} & 0.972 & 0.796 & 0.678 & 0.917 & {\bf 1.000} & 0.429 & 0.552 & 0.945 & {\bf 0.956}  \\
    \bottomrule
    \end{tabular}
    }
    \caption{Performance of detection methods against strong attacks.}
    \label{tab:det}
\end{table}

We report area under the curve (AUC) of the ROC curve as the performance evaluation criterion, as well as the true positive rates (TPR) by thresholding false positive rates (FPR) at 0.01, 0.05 and 0.1, as it is practical to keep mis-classified natural data at a low proportion.  Note that TPR represents the proportion of adversarial examples correctly classified as adversarial, while FPR represents the proportion of natural data incorrectly mis-classified as adversarial. Before calculating the performance metrics, all the samples that can be classified correctly by the model are removed. The results are reported in Table~\ref{tab:det}. ODD~\citep{roth2019odds} and ReBeL~\citep{raghuram2020detecting} show superior or comparable performance over the other detection methods across the two benchmark datasets. 


\section{Concluding Remarks}
\label{sec:conclude}
In this paper we described the attack and defense problems in classification.  We provided a taxonomy of these methods, explained the main ideas behind them, and numerically compared their performances.  As mentioned before, it is our hope that this paper can spark statisticians' interest in this exciting area of research, as we firmly believe that statisticians are well suited to provide unique and significant contributions to the field.  One successful example is the EAD attack of \cite{chen2018ead}, where the elastic-net penalty of~\cite{zou2005regularization} was a key component of the method.  Another successful example is the ER-Classifer of \cite{li2018optimal}, where the method of \cite{levina2005maximum} was used to estimate the intrinsic dimension of the data, and achieve dimension reduction. We conclude this paper with some future work directions that are particularly suited for statisticians to tackle.

\paragraph{Statistical properties of adversarial examples.} It is important and useful to study the statistical properties of adversarial examples, as these can provide useful insights and guidelines for building stronger attack and defense methods.  So far work in this area is rather sparse.  Some notable exceptions include the study of the hidden layer output distribution of adversarial examples~\citep{lee2018simple}, the variance of the hidden output~\citep{raghuram2020detecting}, and the input space feature attribution~\citep{yang2020ml}. 

\paragraph{Efficient modeling for improved decision-based attacks.} 
In reality, many classification systems are blackboxes to the outsiders, and therefore gradient-based and score-based attack methods cannot be applied.  Also, these classification systems typically limit the number of queries any outsider can make, and hence many of the current decision-based attack methods will fail.  The problem of developing a strong decision-based attack method can be viewed as estimating (or modeling) a black-box classification model with a limited number of carefully chosen {\color{black}``design points''} (most likely in a sequential manner) as input to the classifier.



\paragraph{Theoretical guarantee for robust defense methods.} 
With the ever increasing size and complexity of datasets and applications, there will always be a demand for more reliable and efficient defense methods that are scalable.  So far the trend is that many defense methods were shown to have excellent empirical properties at the time of their creation, but then very often they were soon beaten by newer attack methods~\citep{athalye2018obfuscated}.  This suggests that defense methods cannot be blindly trusted.  One ambitious project is to formally define the notion of {\em robustness} in this attack and defense context, which includes attaching a {\em robust score} to any defense method; i.e., the higher the score, the more reliable the method.  This is similar to the use of breakdown point in classical robust statistics.  It is likely that there will be different reasonable definitions for robustness, each designed for one type of attacks.  If a new method achieves high scores for different robustness definitions, then we have strong confidence that this method will work well.


\subsection*{Acknowledgment}
The authors are most grateful to the reviewers, the associate editor and the editor for their most helpful comments.  
The work of Hsieh was partially supported by the National Science Foundation under grants CCF-1934568, IIS-1901527 and IIS-2008173.
The work of Lee was partially supported by the National Science Foundation under grants CCF-1934568, DMS-1811405, DMS-1811661, DMS-1916125 and DMS-2113605.

\bibliographystyle{rss}
\bibliography{main}
\end{document}